\documentclass[11pt]{article}
\usepackage{amssymb,amsmath,amsfonts,geometry,ulem,graphicx,caption,color,setspace,comment,footmisc,caption,natbib,pdflscape,subfigure,array,booktabs,amsthm,mathtools,bibentry,multirow,bbm,bm,afterpage,xr,enumitem}
\usepackage{tabularray}
\usepackage[hidelinks]{hyperref}
\usepackage[labelfont=bf]{caption}
\usepackage{float}
\newpage
\newcommand{\E}{\mathrm{E}}
\newcommand{\Var}{\mathrm{Var}}

\title{The Emergence of Strategic Reasoning of Large Language Models\thanks{We are grateful to  
Soumen Banerjee and Syngjoo Choi for their valuable comments and suggestions. We also thank the participants at the 11th Annual Behavioural Game Theory Workshop (held online) at University of East Anglia. }}
\author{Gavin Kader\thanks{Southwestern University of Finance and Economics. E-mail: gav.kader@outlook.com} \and Dongwoo Lee\thanks{(Corresponding author) Southwestern University of Finance and Economics. E-mail: dwlee05@gmail.com}}
\date{\today}

\usepackage[most]{tcolorbox}

\newtcolorbox{Box1}[2][]{
    colback=black!5, 
    colframe=black, 
    fonttitle=\bfseries, 
    colbacktitle=white, 
    coltitle=black, 
    enhanced,
    breakable, 
    attach boxed title to top left={yshift=-2mm}, 
    boxed title style={
        size=small, 
        boxsep=5pt, 
        width=expand to fit, 
        enlarge left by=4pt, 
        enlarge right by=4pt
    },
    title={#2} #1,
    before upper={\sffamily} 
}

\newtcolorbox{Box2}{
    colback=black!5, 
    colframe=black, 
    enhanced,
    breakable, 
    before upper={\sffamily} 
}

\begin{document}

\newcommand{\gptthree}{GPT-3.5 }
\newcommand{\gptfour}{GPT-4 }
\newcommand{\gptoone}{GPT-o1 }
\newcommand{\claudeone}{Cl-1 }
\newcommand{\claudetwo}{Cl-2 }
\newcommand{\claudethree}{Cl-3S }
\newcommand{\claudefour}{Cl-4ST }
\newcommand{\geminione}{Ge-1 }
\newcommand{\geminitwo}{Ge-2T }
\newcommand{\gptthreens}{GPT-3.5}
\newcommand{\gptfourns}{GPT-4}
\newcommand{\gptoonens}{GPT-o1}
\newcommand{\claudeonens}{Cl-1}
\newcommand{\claudetwons}{Cl-2}
\newcommand{\claudethreens}{Cl-3S}
\newcommand{\claudefourns}{Cl-4ST}
\newcommand{\geminionens}{Ge-1.5}
\newcommand{\geminitwons}{Ge-2T}
\begin{titlepage}
\maketitle

\begin{abstract}
    
\noindent As large language models (LLMs) have demonstrated strong reasoning abilities in structured tasks (e.g., coding and mathematics), we explore whether these abilities extend to strategic multi-agent environments. We investigate strategic reasoning capabilities -- the process of choosing an optimal course of action by predicting and adapting to others’ actions -- of LLMs by analyzing their performance in three classical games from behavioral economics. Using hierarchical models of bounded rationality, we evaluate three standard LLMs (ChatGPT-4, Claude-3.5-Sonnet, Gemini 1.5) and three reasoning LLMs (OpenAI-o1, Claude-4-Sonnet-Thinking, Gemini Flash Thinking 2.0). Our results show that reasoning LLMs exhibit superior strategic reasoning compared to standard LLMs (which do not demonstrate substantial capabilities) and often match or exceed human performance; this represents the first and thus most fundamental transition in strategic reasoning capabilities documented in LLMs. Since strategic reasoning is fundamental to future AI systems (including Agentic AI), our findings demonstrate the importance of dedicated reasoning capabilities in achieving effective strategic reasoning.

\medskip
\noindent\textbf{Keywords:} Strategic Reasoning, Large Language Models, Level-$k$ Model, Cognitive Hierarchy Model

\end{abstract}

\end{titlepage}

\section{Introduction}\label{sec:introduction} 
As the reliance on Large Language Models (LLMs) is now pervasive, it is vital to assess whether the mechanisms through which LLMs reason are also capable of \textit{strategic} reasoning -- the process of choosing an optimal course of action by predicting and adapting to the actions of others in multi-agent environments. While research has shown LLMs' reasoning skills in individual decision-making (economic rationality) \citep{chen_emergence_2023,kim_learning_2024}, mathematics \citep{ahn_large_2024, zhou_mathattack_2024}, simple reasoning problems \citep{webb_emergent_2023, hagendorff_human-like_2023}, and coding tasks \citep{openai_openai_2024, anthropic_claude_2024}, this does not inherently imply they possess strategic reasoning capabilities. In fact, strategic reasoning is an emergent capability virtually absent in standard LLMs, only manifesting with the advent of reasoning LLMs, representing the most fundamental and sharpest transition in strategic reasoning capabilities since the development of LLMs.\footnote{While earlier standard LLMs trained before the first reasoning LLMs demonstrate a lack of inherent strategic reasoning, standard LLMs developed later are not incapable of strategic reasoning. After the introduction of reasoning LLMs, standard LLMs are likely to be exposed to more reasoning-based thinking and tasks (or even fine-tuned or undergone a form of knowledge distillation from reasoning LLMs) allowing standard LLMs to develop higher-order strategic reasoning in a way that is crudely analogous to reasoning LLMs.} 

Such capabilities are crucial for the increasingly popular use of Agentic AI (AAI), which involves multi-agent decision-making environments with AI/LLM systems having competitive or cooperative objectives that require constant anticipation and adaptation for autonomous, real-time tasks. Key real-world applications of AAI include autonomous financial agents that anticipate market movements to make trading decisions \citep{luo_agent_2025,an_finverse_2024}, supply chain systems that coordinate resources by predicting demand \citep{almutairi_resilient_2025,xu_implementing_2024}, and negotiation systems requiring agents to infer and strategically anticipate responses \citep{abdelnabi_cooperation_2024,kwon_are_2024}. Unlike individual decision-making, these types of applications require agents to form beliefs about how others (LLMs and/or people) will act in order to complete tasks.\footnote{More generally, achieving Artificial General Intelligence (AGI) - AI that performs a wide range of cognitive tasks at, or beyond, human-level proficiency \citep{singh_enhancing_2024} - necessitates systems capable of strategic reasoning, making it essential to benchmark these capabilities \citep{mumuni_large_2025}.} The strategic sophistication of these AI agents therefore has consequences for human decision-makers. As people increasingly use LLMs for strategic advice (negotiations, investments, etc.), more sophisticated LLMs may provide better outputs for improved decision-making in strategic contexts. From a market perspective, variations in strategic sophistication across LLMs create potential asymmetries where users of less sophisticated LLMs may be exploitable by those employing more strategically sophisticated systems. Thus, investigating the strategic sophistication of LLMs has broader implications beyond benchmarking model performance.

In this study, our objective is to explore this emergence of strategic reasoning capabilities \citep{wei_emergent_2022,schaeffer_are_2023} leveraging OpenAI's ChatGPT-4 and OpenAI-o1-preview (\gptfourns, \gptoonens), Anthropic's Claude-3.5-Sonnet and Claude-4-Sonnet-Thinking (\claudethreens, \claudefourns), and Google's Gemini 1.5 Flash and Gemini Flash Thinking 2.0 Experimental (\geminionens, \geminitwons). We investigate whether reasoning LLMs, specifically designed with reasoning techniques (\gptoonens, \claudefourns, and \geminitwons) exhibit sophisticated strategic reasoning abilities, and whether such abilities were already present in more standard LLMs. To evaluate and deepen our understanding of LLMs’ strategic reasoning, we adopt a behavioral economics framework -- a ubiquitous approach in studying human behavior -- to provide valuable insights into how well LLMs can navigate strategic environments.

We explore strategic reasoning using three classical games from the behavioral economics literature: the p-Beauty Contest Game (pBCG), the Guessing Game (GG), and the 11-20 Money Request Game (MRG) \citep[][respectively]{nagel_unraveling_1995,costa-gomes_cognition_2006,arad_1120_2012}, allowing us to evaluate LLMs by estimating their depth of strategic reasoning and investigate whether we observe similarities/differences across games.\footnote{In human subjects, \cite{georganas_persistence_2015} show that strategic reasoning can differ across games.}$^{,}$\footnote{Using behavioral games (dictator game, ultimatum game, trust game, bomb risk game, and public goods game), \cite{mei_turing_2023} show that LLMs exhibit behavioral traits such as trust, fairness, risk-aversion, altruism, and cooperation, with responses that are statistically indistinguishable from human behavior.} These games span distinct theoretical demands. In the pBCG, iterated best-response reasoning leads to a unique symmetric equilibrium, so equilibrium actions are identical between players. By contrast, the GG features role-specific incentives and yields asymmetric equilibrium actions between players. Finally, the MRG requires probabilistic reasoning: an equilibrium is a randomization of actions rather than a single deterministic action for players. Using such games provides us with essential tools to evaluate LLMs, as the focus is on logic and strategic reasoning through clear, well-defined rules and, vitally, action spaces that allow us to directly test the depth of strategic reasoning abilities. 

Two of the most profound models from behavioral economics that capture the recursive process involved in limited strategic sophistication are the level-$k$ model \citep{stahl_experimental_1994,stahl_players_1995,nagel_unraveling_1995} and the cognitive hierarchy (CH) model \citep{camerer_cognitive_2004}.\footnote{Abundant studies in behavioral economics have shown that the Nash equilibrium, derived from the assumption of full rationality, does not match well with observed play \citep{camerer_advances_2004, camerer2011behavioral, bernheim_handbook_2019}. This has given rise to models of limited strategic reasoning.} These alternative models of bounded rationality have been able to provide more accurate predictions of behavior. Studies in behavioral economics have largely focused on human adults with some research done with children \citep{brocas_evolution_2020, hermes_if_2022,crawford_structural_2013}. We apply both models to evaluate LLMs by measuring their level of strategic reasoning represented by a hierarchy of iterated reasoning -- an investigation which is yet to be done with LLMs. 

The \textit{lowest rank} in the hierarchy, say rank-$0$, is completely non-strategic, that is, actions are made without taking into account the actions of others.\footnote{A common assumption is that rank-$0$ behavior is consistent with choosing an action as if drawing uniformly from the action space.} The next rank represents more strategic behavior (rank-$1$) with beliefs that all others are non-strategic (rank-$0$). Similarly, rank-$2$ beliefs are that all others consist of lower ranks. In the level-$k$ model, these lower ranks, referred to as levels, are always exactly one level lower of reasoning, e.g., level-2 ($L_2$) assumes that all others are level-1 ($L_1$). In the CH model, these lower ranks, referred to as steps, accommodate a combination of all lower steps of reasoning, e.g., step-2 assumes that all others are a mixture of step-$0$ and step-$1$. As we move up the hierarchy, this pattern continues for higher ranks. In this process, knowing how many stages of iterated reasoning the others are utilizing (i.e., the others' levels) is essential. In this way, we aim to answer to what extent are LLMs strategically sophisticated.

To evaluate the strategic reasoning of LLMs, we instruct LLMs to play the different games under various treatments. To obtain a distribution of responses from each LLM, we request 100 responses, with each prompt being independent -- akin to conducting a laboratory experiment with 100 subjects. Additional treatments include varying the parameters of the games themselves.\footnote{The temperature parameter of LLMs can also influence output in terms of randomness and creativity. We use a baseline temperature of 0.5 and also test a high (0.75) and low (0.25) temperature. Section \ref{sec:temp} of Appendix B contains the results of the high and low temperatures with overarching takeaway that temperature has little effect on observed strategic reasoning.} By estimating the level-$k$ model and the CH model, we are able to determine the depth of reasoning (up to a fixed level/step) by matching the theoretical predictions of each model for each game with the observed data. Our overall results are (i) reasoning LLMs demonstrate higher strategic reasoning compared to standard LLMs, (ii) with sufficient feedback, LLMs are able to learn and achieve higher orders of strategic reasoning (especially standard LLMs), (iii) reasoning LLMs learn faster than standard LLMs, (iv) standard LLMs consistently exhibited worse strategic reasoning than typical human subjects (from the experimental economics literature), and naturally implying (v) reasoning LLMs often exhibit higher strategic reasoning than human subjects (generally in favor of \gptoone over \claudefour and \geminitwons).  

The enhanced reasoning techniques of the reasoning LLMs is the most likely explanation for their higher strategic reasoning. Specifically, \gptoonens, \claudefourns, and \geminitwo leverage specialized large-scale reinforcement learning to enhance their reasoning skills, with chain-of-thought reasoning -- a process where the model tackles tasks step-by-step, approaching problems incrementally \citep{wei_chain--thought_2023}, which, in practice, means training LLMs to break down complex tasks into smaller, manageable steps, mirroring human-like reasoning.\footnote{As opposed to just prompting them to use chain-of-thought reasoning.} With this approach, reasoning LLMs refine their reasoning by exploring various strategies and recognizing potential mistakes, further enhancing their problem-solving capabilities in an iterative, and (seemingly) reflective process, ensuring more careful evaluation before responding.\footnote{OpenAI-o1-preview is the first model of this type and therefore serves as an essential benchmark.}

Using standard methodologies and models from behavioral economics, and experimental economics, we are able to concretely evaluate the strategic reasoning capabilities of LLMs using models of bounded rationality. We contribute to the ongoing interest in comparing between LLMs as well as comparisons with human behavior. Further, we highlight that with sufficient feedback, even LLMs which exhibited poor performance in one-shot games can refine their strategies, which draws a parallel with human subjects \citep{nagel_unraveling_1995,ho_iterated_1998,weber_learning_2003} and the reasoning LLMs. By incorporating thinking/reasoning skills into the mechanism through which LLMs operate, it gives a clear inherent indication as to why reasoning LLMs perform vastly better than other LLMs and are capable of outperforming human subjects. Our study emphasizes the importance of continuous improvement in these mechanisms, particularly for interactive tasks requiring higher-order reasoning.

\section{p-Beauty Contest Game}
We begin with an overview of the pBCG, examining its theoretical predictions from behavioral game theory, which accounts for both full and bounded rationality. After briefly describing how these models are estimated, we explain our approach to implementing the pBCG with LLMs.

\medskip
\noindent\textbf{Game.} A group of $n\geq2$ players simultaneously choose a number in the interval $[0,100]$. The winner of the game is the player whose number is closest to a given target multiplier ($p$) of the average number chosen by all players.

\medskip
\noindent\textbf{Theoretical Predictions.} Nash equilibrium requires no player has an incentive to deviate from a given strategy profile, assuming all other players' strategies remain the same. The equilibrium is based on the assumption of common knowledge of rationality that all players are rational and that is known to all players. In the pBCG with $p<1$, a unique Nash equilibrium is that all players choose 0 as no player gains by choosing another number when all others choose 0. Similarly, in the pBCG with $p>1$, a unique Nash equilibrium is that all players choose 100. 

Both in the level-\textit{k} and the cognitive hierarchy (CH) models, it is assumed that players are boundedly rational in the sense that each player deems their level/step of reasoning is higher than that of others. Specifically, the level-\textit{k} model assumes that a player believes that all others are one level lower in their reasoning compared to themselves. The CH model assumes that a player believes that the reasoning steps of the others are distributed (strictly) below her own, according to a Poisson distribution. For both models, we use the concept of \textit{best guess} which is a target number based on an average guess of numbers from other players and the target multiplier ($p$).\footnote{In game theory, the concept of a \textit{best response} is often employed to derive an equilibrium -- a set of strategies that provide the most favorable outcome based on a belief about the strategies of others. In games such as these in which a best response given a belief may not be uniquely defined, we use a best guess as a reasonable best response, common to the behavioral economics literature, making it possible to identify a level/step of reasoning.}

In the level-\textit{k} model, the best guess for each reasoning level, denoted by $l_k$ for $k=0,1,\ldots$, is defined as:
\begin{equation*}
    l_0=50,~~l_1=l_0\times p=50p,~~l_2=l_1\times p=50p^2,\ldots,~~l_k=l_{k-1}\times p=50p^{k},
\end{equation*}
where an $L_0$ type chooses $l_0=50$, which is the average guess when choosing uniformly at random from $[0,100]$. A $L_k$ type, $k=1,2,\ldots$, believes that all other players are $L_{k-1}$, choosing $l_{k-1}=50\times p^{k-1}$. Clearly, $l_k$ converges to $0$ and $100$ for $p<1$ and $p>1$, respectively, in the interval $[0,100]$ when the reasoning level $k$ goes to infinity.

The CH model assumes that each player is playing against a distribution of players who are using lower reasoning steps. This distribution is given by a Poisson distribution $f(\cdot;\tau)$, where $\tau$ is the mean (and the variance) of the number of reasoning steps. Specifically, a step-$k$ type believes that they are playing against a distribution of players with reasoning steps from $0$ to $k-1$. This distribution is given by $f_k(j;\tau)$, $j=0,1,\ldots,k-1$, such that
\begin{equation}
    f_k(j;\tau)= \left(\frac{e^{-\tau}\tau^j}{j!}\right) \Big\slash \left(\sum_{m=0}^{k-1} \frac{e^{-\tau}\tau ^m}{m!} \right),\label{conditional pdf}
\end{equation}
where $f_k(j;\tau)$ is the conditional probability for $j=0,1,\ldots,k-1$. 

The best guess for reasoning steps, denoted by $s_k$ for $k=1,2,\ldots$, is defined as
\begin{equation*}
    s_k = \sum_{j=0}^{k-1} f_k(j;\tau)s_j,
\end{equation*}
where the average guess for step-0 is $s_0=50$, that is, a random choice from the interval $[0,100]$. A step-$k$ type believes that all others are step-$0$ to step-$(k-1)$, according to the Poisson probability density $f_k(j;\tau)$. For example, with $\tau=1.5$, a step-2 player believes that they are playing against a combination of step-0 and step-1 players, with the proportions of each being 40\% and 60\%, respectively. Given these proportions, a step-2 player would choose to play $p$ times the expected response of the other players.

\medskip

\noindent\textbf{Estimation.} From these theoretical predictions, we estimate the parameters for both the level-$k$ and CH models by maximum likelihood estimation (MLE). For the level-$k$ model, we estimate the proportions of responses corresponding to each level from $L_0$, $L_1$, $L_2$, $L_3$, $L_4$, and $L_\infty$ (Nash) predictions. We assume that each response is a best guess with (independent and identically distributed) noise. Estimation of the level-$k$ model gives the distribution of responses according to the different levels of reasoning. For example, if responses are concentrated around the lower-level types' guesses, compared to higher-level types' guesses, then overall responses from the former must exhibit less strategic reasoning than the latter. The estimated level-$k$ model will reflect this in estimating higher frequencies for the lower levels of reasoning. Similarly for the CH model, we estimate the distribution of step-types using a Poisson distribution (fully parameterized by $\tau$), as defined above, with a similarly defined noisy best guess based on the (conditional) distribution of lower step-types. For example, a higher/lower estimated $\tau$ is indicative of higher/lower level reasoning, on average, as this indicates more/fewer (average) steps of reasoning.\footnote{For example, in pBCG, a $\hat\tau = 0$ implies that all players are step-$0$ types, whereas $\hat\tau \rightarrow \infty$ indicates players are fully rational, as \textit{infinitely} iterated elimination of dominated strategies leads to a unique Nash equilibrium.  A $\hat\tau = 1.5$ indicates that the average step-type is 1.5. Further, as the frequency of types $f(k) $ is Poisson-distributed, the ratio $ \frac{f(k-1)}{f(k-2)} $  favors step-$(k-1)$ types when $ \tau $ is large, i.e., step-$k$ types play as if there are predominantly step-$(k-1)$ type when $ \tau $ is large.} A detailed description of the estimation for both models is provided in Section \ref{est:pBCG} of Appendix B.

\medskip

\noindent\textbf{Instructing LLMs.} To collect data, we use the respective application programming interfaces (API) with Python/R from OpenAI, Anthropic, and Google. The exact input prompts for the understanding questions and the tasks are provided in Sections \ref{sec:understanding} and \ref{sec:tasks} of Appendix B.

\subsection{Design}
The baseline condition refers to a specific set of parameters of the pBCG.

\noindent \textbf{Baseline condition:} The total number of participants is 11 ($n=11$), the target statistic is the average, and the target proportion is $2/3$ ($p=2/3$). This represents a classical pBCG with the most commonly used parameters in the literature. We ask each LLM to complete the baseline pBCG 100 times, providing us with a distribution of responses.

\medskip

\noindent \textbf{Alternative conditions:}  To provide comparisons with the baseline condition, we change one of the parameters while keeping the others fixed. LLM responses are collected for (i) $p\in\{ 1/2, 4/3\}$, (ii) $n\in\{2,\text{unspecified}\}$, (iii) $\{\text{median}\}$. Varying these parameters does not change the substance of the baseline pBCG, but offers additional insights into LLM strategic reasoning. 

\medskip

\noindent\textbf{Multiple rounds:} We collect responses from a repeated version of the pBCG, with feedback, conducted over 10 rounds, using the same parameters as the baseline condition ($n=11$), including $p=\{4/3\}$. Each of the LLMs play within a group of 10 other LLM subjects of the same model. After each round, each player receives information about the average bid and the target number ($p \times$average). Additionally, players are privately informed of their win (or loss) but not the identity of the winner nor the winning bid (following \cite{ho_iterated_1998}). Players are provided with the cumulative history of the game from round 1 to 10 within their prompts.

\medskip

\subsection{Baseline Condition Results}

Before proceeding with data collection, we conducted a series of preliminary checks involving several questions for whether LLMs  understand the game-specific rules and winning conditions -- and whether they choose a best response when given the actions of the other players to confirm their ability to respond optimally. We asked each LLM to answer each question 25 times. Each LLM consistently answers all of the questions related to the rules of the game, as well as when winning conditions were satisfied (or not) correctly. All LLMs also provided appropriate answers for a best response in the baseline condition.

\begin{figure}[ht!]
    \centering
    \includegraphics[width=1\textwidth]{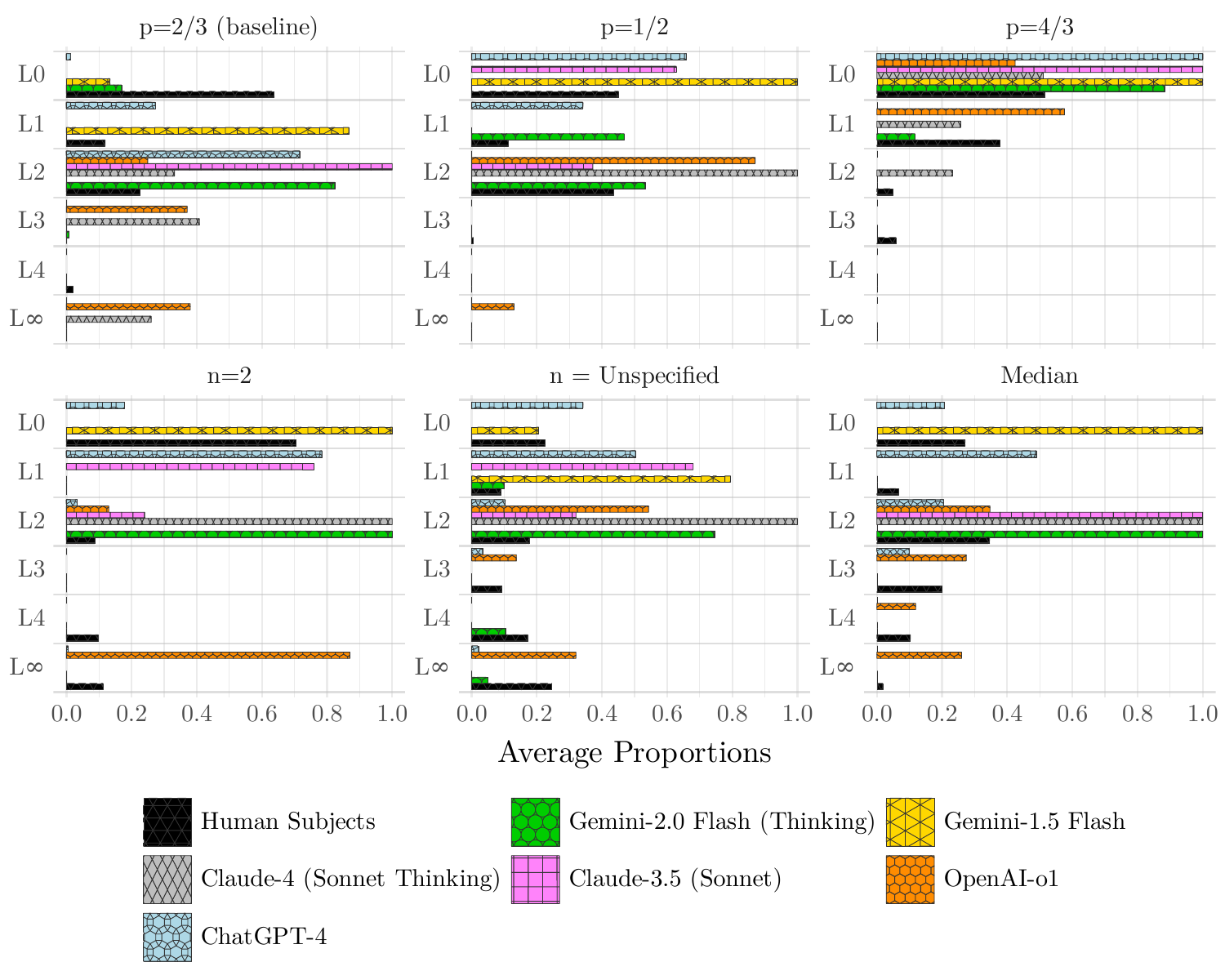}
    \caption{Estimates of the proportions of level-$k$ types for the pBCG}
    \label{fig:lk_pb}
\end{figure}

The reasoning LLMs are more strategically sophisticated than the standard LLMs by at least 1 to 2 levels of thinking on average, with \gptoone exhibiting the highest level of strategic reasoning overall, summarized in Figure \ref{fig:lk_pb}.\footnote{All non-zero proportions in Figure \ref{fig:lk_pb} are statistically significant at the 5\% level with inference based on randomized resampling (bootstrap).} As is well known in the experimental economics literature \citep[see][as representative examples]{nagel_unraveling_1995,bosch-domenech_one_2002}, human subjects are typically a combination of $L_0$, $L_1$, $L_2$, and $L_\infty$, with $L_\infty$ types being rare. To draw direct comparisons with human subjects, we also do estimation using data from \cite{nagel_unraveling_1995} (different $p$'s), \cite{grosskopf_two-person_2008} ($n=2$), \cite{bosch-domenech_one_2002} ($n=\text{unspecified}$), and \cite{hermes_if_2022} (median). From Figure \ref{fig:lk_pb}, we observe that the distribution of levels for \gptoonens, \claudefourns, \geminitwons, and \claudethree are shifted toward higher levels of strategic reasoning compared to human subjects. Overall, \gptoone and \claudefour exhibit the highest levels of strategic reasoning by large margins.\footnote{Using data from \cite{nagel_dataset_1995} ($N=67$), we also conduct Kolmogorov-Smirnov (KS) tests for the distribution of raw results. Results are summarized in Table \ref{tab:pbcg_tests}. Focusing on those cases in which exactly one of the one-sided tests is rejected, one-sided KS tests suggest that \gptfourns, \gptoonens, \claudefourns, and \geminitwo outperform human subjects at any reasonable significance levels. Two-sided tests suggest the distributions between LLMs and human subjects are all statistically different from each other at any reasonable significance level. Data made available at \citep{nagel_dataset_2002}.}

Table \ref{tab:ch_pb} shows estimates for $\tau$ alongside respective 95\% confidence intervals.\footnote{Confidence intervals are estimated from randomized resampling (bootstrap). This is true for all tables with estimates from the CH model in this paper.}  On average, \geminione exhibits just under 1-step of strategic reasoning with estimates not dissimilar to human subjects (see Table \ref{tab:pbcg_tests}). \gptoonens, and \claudefour vastly outperformed the other LLMs, and exhibit higher levels of strategic reasoning than human subjects, with \gptoone marginally exhibiting the highest average steps of reasoning, corroborating with the estimates from the level-$k$ model.

\begin{table}[]
\centering
\resizebox{\textwidth}{!}{%
\begin{tabular}{l|>{\centering\arraybackslash}m{2.5cm}|>{\centering\arraybackslash}m{2.5cm}|>{\centering\arraybackslash}m{2.5cm}|>{\centering\arraybackslash}m{2.5cm}|>{\centering\arraybackslash}m{2.5cm}|>{\centering\arraybackslash}m{2.5cm}}
\toprule
 & $p=2/3$ (baseline) & $p=1/2$ & $p=4/3$ & $n=2$ & $n=\text{unspecified}$ & \text{Median} \\ \midrule
\multirow{2}{*}{\gptfour}     & 2.39            & 0.56            & 0.00            & 0.87            & 0.83            & 1.09        \\ 
                              & (2.23,2.46)     & (0.56,0.79)     & (0.00,0.00)     & (0.79,0.94)     & (0.69,1.01)     & (0.83,1.36) \\ \midrule
\multirow{2}{*}{\gptoone}     & 4.42            & 2.52            & 0.95            & 7.09            & 3.43            & 4.00        \\ 
                              & (4.00,4.77)     & (2.52,2.99)     & (0.744,0.96)    & (6.42,8.00)     & (3.11,3.76)     & (0.00,4.86) \\ \midrule
\multirow{2}{*}{\claudethree} & 2.87            & 1.50            & 0.00            & 1.17            & 1.26            & 2.87        \\ 
                              & (2.87,2.87)     & (1.31,1.57)     & (0.00,0.00)     & (1.12,1.24)     & (1.18,1.34)     & (2.65,2.88) \\ \midrule
\multirow{2}{*}{\claudefour}  & 4.00            & 2.83            & 0.00            & 2.33            & 2.15            & 2.45        \\ 
                              & (0.00,4.23)     & (2.70,2.99)     & (0.00,0.00)     & (2.24,2.42)     & (2.10,2.21)     & (2.39,2.54) \\ \midrule
\multirow{2}{*}{\geminione}   &  0.86           &  0.00           &  0.00           &  0.00           &  0.77           &  0.00       \\ 
                              &  (0.79,0.91)    & (0.00,0.00)     & (0.00,0.00)     & (0.00,0.00)     &  (0.70,0.84)    & (0.00,0.00)  \\ \midrule
\multirow{2}{*}{\geminitwo}   &  2.46           &  2.12           &  0.00           &   1.86          &  2.22           &  2.46        \\ 
                              &  (2.39,2.72)    &  (1.99,2.12)    & (0.00,0.00)     &  (1.79,1.91)    &   (2.01,2.41)   &  (2.46,2.92) \\ 
                              \midrule
\multirow{2}{*}{Human Subjects}   &  0.00           &  1.08           &  0.36           &   0.00          &  3.01           &  2.46        \\ 
                              &  (0.00,1.09)    &  (0.00,2.14)    & (0.00,0.57)     &  (0.00,0.00)    &   (2.91,3.12)   &  (0.00,2.46) \\ 
                              \bottomrule
\end{tabular}%
}
\caption{Estimates of $\tau$ from the CH model for different pBCG parameters. 95\% confidence intervals (bootstrapped) are shown in parentheses.}
\label{tab:ch_pb}
\end{table}

\subsection{Alternate Conditions Results}
As with the baseline condition, we check whether LLMs understand the game (specific rules \& win condition, and best responses). All understanding questions were answered correctly by each LLM.

From Figure \ref{fig:lk_pb}, \geminione is consistent with their baseline estimates and the $n=\{\text{unspecified}\}$ condition but shows a lack of strategic reasoning in the remaining conditions, exhibiting overall drops in strategic reasoning from baseline. \claudethree only manages to stay at $L_2$ for the median condition with drops to combinations of $L_0$/$L_1$ and small proportions of $L_2$ behavior. Only \gptoone and (marginally) \geminitwo exhibit $L_\infty$ behavior. This is particularly true when $n=\{2\}$, where only \gptoone is virtually consistent with a fully rational player.\footnote{Increases in strategic reasoning are not surprising in this condition, given that the game is less complicated when there are fewer players, but with dramatic/modest improvements for \gptoonens/\geminitwons.}  Overall, for all conditions except when $p=\{4/3\}$, only the reasoning LLMs exhibit significant $L_2$+ behavior, although with \claudefour showing significant drops compared to baseline, and with \gptoone exhibiting the highest overall strategic reasoning.

When $p=\{4/3\}$, only the reasoning LLMs exhibit any non-random behavior. Despite this, compared to their baselines, reasoning LLMs drop drastically in their observed strategic reasoning. A plausible explanation for poor performance in this condition is that LLMs are vastly trained on pBCGs which involve iterating downward to a lower number (i.e., every other condition) as the literature mainly focuses on these versions of pBCG. This lack of flexibility means it may be difficult to adapt to a condition which involves a different iterative process, albeit similar for a human.

When compared with human subjects, the standard LLMs are typically dominated in strategic reasoning for all the alternate conditions. \claudethree does comparably when $p=\{1/2\}$ against human subjects but is dominated for all other conditions. Both \claudefourns, and \geminitwo perform better and are comparable or better in all conditions except the $n = \text{unspecified}$ condition. \gptoone outperforms human subjects in terms of strategic reasoning in all alternate conditions, except when $p=\{4/3\}$ in which case it is slightly outperformed by human subjects, yet still comparable in reasoning.\footnote{We conduct similar KS tests as was done for the baseline condition to compare raw responses. Results are summarized in Table \ref{tab:pbcg_tests}. We focus on cases in which exactly one of the one-sided tests is rejected. For the target proportion conditions ($N_{p=1/2}=48;N_{p=4/3}=51$), all LLMs, excluding \claudefourns, are outperformed by human subjects with \geminitwo outperforming when $p=\{1/2\}$  \citep{nagel_unraveling_1995}. \gptoone outperforms when $n=\{\text{unspecified}\}$ ($N=120$) and the median ($N=1468$) conditions \citep{hermes_if_2022,bosch-domenech_one_2002}. For the $n=\{2\}$ alternate condition in \citep{grosskopf_two-person_2008} ($N=132$), only \gptoone clearly outperforms human subjects.}

The estimates of the CH model corroborate the finding of low strategic reasoning across all alternate conditions for standard LLMs and that all reasoning LLMs (weakly) outperform standard LLMs, summarized by Table \ref{tab:ch_pb}. When $p=\{4/3\}$, all LLMs perform poorly, with \gptoone barely outperforming the other LLMs.\footnote{This is consistent with \cite{ho_iterated_1998} who also estimate $\tau=0$ when $p=1.3$ ($>1$).} From Table \ref{tab:ch_pb} we also observe that human subjects overall outperform standard LLMs in strategic reasoning, except when $n=2$ where human subjects perform surprisingly poorly, as discussed in \cite{grosskopf_two-person_2008}. \claudethree is comparable with human subjects in most conditions except the $n=\{\text{unspecified}\}$ in which human subjects exhibit almost 2 more steps of reasoning. \geminitwo is comparable in the median condition, and outperforms when $p=\{1/2\}$, but outperformed otherwise. \gptoone vastly outperforms human subject in all treatments in terms of strategic reasoning, by almost 2 steps of thinking on average, with only the $p=\{4/3\}$ being comparable with human subjects.

\medskip

\subsection{Multiple Rounds Results} 
From the previous analysis, it is evident that the majority of LLMs display limited strategic reasoning in one-shot pBCG, particularly the standard LLMs. To investigate further, we explore a repeated version of pBCG with multiple rounds to see whether LLMs can learn to adjust their responses. 

\begin{figure}[ht!]
\centering
    \includegraphics[width=\linewidth]{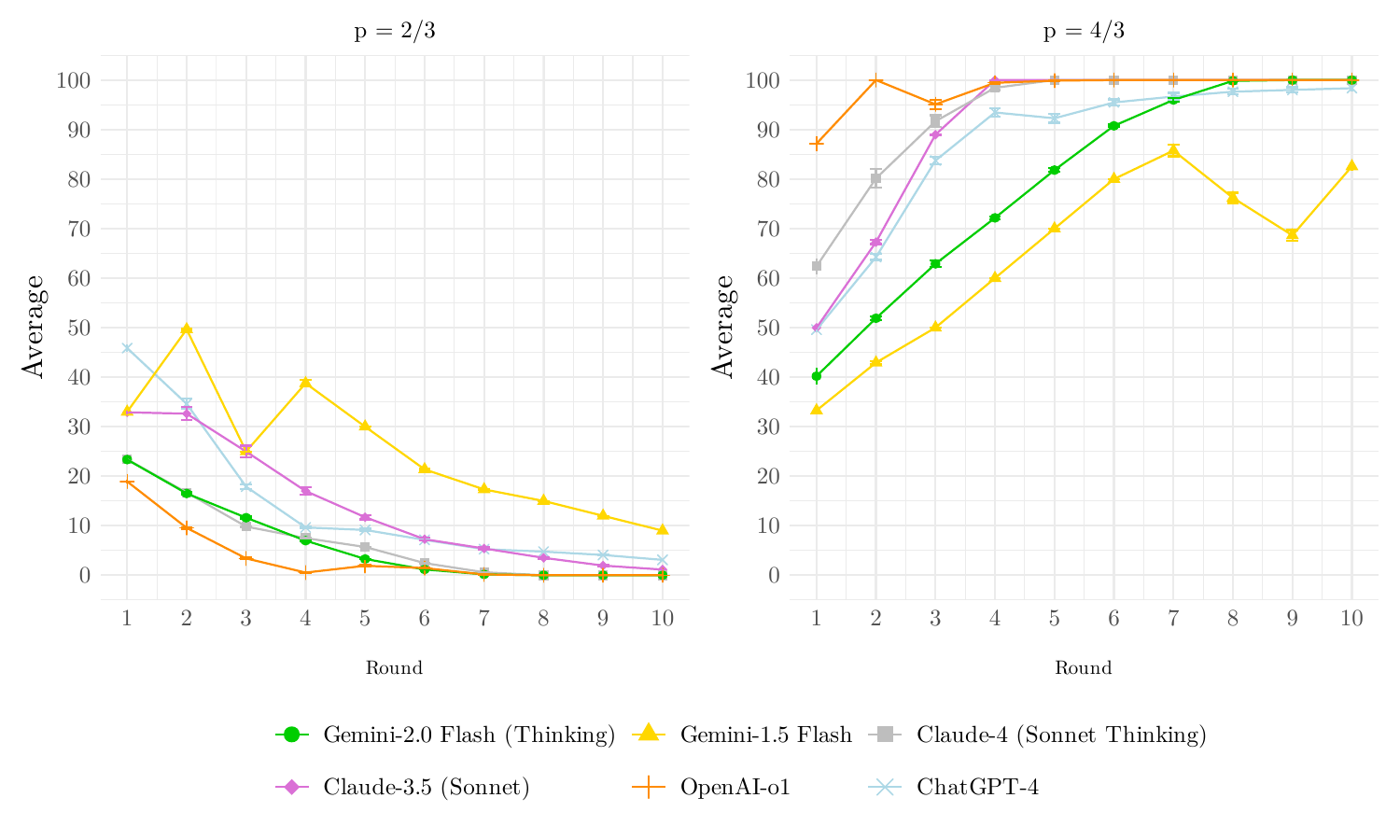}
    \caption{Time series of responses for multiple rounds pBCG. This figure consists of two figures for p = 2/3 and p = 4/3, respectively. Each figure shows the average responses in each round for each LLM. Vertical bars are 95\% confidence intervals}
    \label{fig:feedback}
\end{figure}

From Figure \ref{fig:feedback}, across all LLMs, \gptoonens's initial responses are always closest to the fully rational prediction, approximately exhibiting $L_2$ behavior.  For $p=\{2/3\}$, the initial responses of the other LLMs are consistent with $L_0$ and $L_1$ (or approximate $L_2$ behavior, approximately consistent with their one-shot levels of strategic reasoning. A similar result occurs when $p=\{4/3\}$ with only $L_0$ behavior being observed. 

When $p=\{2/3\}$, all LLMs exhibit a downward trend to the rational prediction; \geminione only manages to converge to approximately $L_3$ behavior, but does not approximate $L_\infty$ behavior unlike the other LLMs which do get to (in the close vicinity of) the rational prediction by round 10. Likewise, for $p=\{4/3\}$, all LLMs show upward trends across all rounds with only \geminione not achieving the rational prediction. In general, the reasoning LLMs achieve the rational prediction faster than the standard LLMs (that converge), suggesting that even standard LLMs are capable of learning and adapting to strategic environments with sufficient feedback. \gptoone clearly outperforms all other LLMs, particularly in rate of convergence, and initial responses.

We are able to draw comparisons and several similarities with human subjects in \cite{ho_iterated_1998} and \cite{weber_learning_2003}. For human subjects, the rate of convergence (when convergence occurs) is much faster when $p=\{4/3\}$ compared to $p=\{2/3\}$.\footnote{This is intuitive as, in principle, it takes fewer levels/steps of reasoning to achieve the rational prediction when $p=\{4/3\}$ compared to $p=\{2/3\}$.} In general, for the reasoning LLMs, the patterns of convergence and the (boundedly rational) responses in the first round are comparable with human subjects in \cite{ho_iterated_1998} when $p=\{2/3\}$. When $p=\{4/3\}$, LLMs seem to converge to $L_\infty$ behavior faster than human subjects. Overall, from their observed behavior, LLMs are capable of learning and adapting in ways that are comparable to human subjects. Only the reasoning LLMs perform consistently better than human subjects, a result which is consistent throughout the entire analysis of pBCG.

\section{Guessing Game}

The Guessing Game (GG) from \cite{costa-gomes_cognition_2006} is a two-player game in which both players make a guess about the opponent's guess, similar to the pBCG. Unlike the pBCG, both players may have different target multipliers and different domains for their answers, which is common knowledge to both players.\footnote{Recall that in pBCG, all players have the same target multiplier -- $1/3$, $1/2$, or $4/3$ -- and the same range of choices, $[0,100]$.} This asymmetry in target multipliers and answer domains leads to an asymmetric rational prediction in Nash equilibrium between both players.  An appealing feature of the GG is its dominance-solvability within a finite number of rounds.\footnote{A game is called \textit{dominance-solvable} if the iterated elimination of dominated strategies results in a unique Nash equilibrium, as in pBCG and GG, but not in MRG.} This can bring an additional advantage as we can easily identify subjects' decision rules, and therefore, types, for both the level-$k$ model, and the CH model, no matter the number of levels/steps of reasoning.\footnote{Recall the pBCG with a choice range of $[0,100]$ is, in principle, dominance solvable only in an infinite number of rounds for targets that are less than one ($p<1$) meaning we may not be able to pin down precisely the exact levels/steps of reasoning required to reach equilibrium. Although, in practice, subjects/LLMs most likely choose the Nash equilibrium response once their steps of reasoning reasonably converges to 0.}$^,$\footnote{From a practical standpoint, the GG is not as commonly studied in the literature, suggesting LLMs will not be trained on many sources involving the GG as compared to the pBCG; this is also true for the MRG, analyzed in the next section.}
 
\medskip
\noindent\textbf{Game.} There are two players, 1 and 2. Player 1 makes a \textit{guess}, $x^1$, within a range defined by a lower limit, $a^1$, and an upper limit, $b^1$. Similarly, for player 2, $x^2 \in [a^2,b^2]$. Each player $i$'s guess should be as close to their target number, which is their target multiplier $p^i$ multiplied by the other player $j$'s guess, that is, $p^i \times x^j$. Specifically, player $i$'s payoff is given by
\begin{equation*}
    \pi^i=\max\{0,200-d^i\}+\max\{0,100-d^i/10\},
\end{equation*}
where $d^i$ is the absolute difference value between player $i$'s guess and player $i$'s target number, defined as $d^i\equiv |x^i(a^i,b^i)-p^i\cdot x^j(a^j,b^j)|$. 
\medskip

The payoff function $\pi^i$ is uniquely maximized when $d^i=0$ for player $i$ and provides an additional incentive for a closer guess $d^i\leq 200$. Hence, unlike the pBCG, each player has a unique best response to every opponent's guess in the GG. 
\medskip

\noindent\textbf{Theoretical Predictions.} 
Predictions under the level-$k$ model are made in a similar way to the pBCG, illustrated with an example. Suppose player 1 has a choice range in $[100,900]$ with $p^1=0.5$ and player 2 has a choice range in $[100,500]$ with $p^2=1.5$. An $L_0$ type randomizes a guess within their choice range. Hence, $l_0^1=(100+900)/2$ and $l_0^2=(100+500)/2$. An $L_1$ type believes that the opponent is a $L_0$ type and best responds to their belief. The best guess for player 1 with $L_1$ is $0.5\times l_0^2 =150$, which is within the range of player 1's choice. Hence, $l_1^1=150$. Similarly, the best guess for player 2 with $L_1$ is $1.5\times l_0^1=750$. As this is not within player 2's choice range, $l_1^2$ is 500 which is the closest number to 750 within $[100,500]$. This process continues, generating predictions for each level of reasoning in the level-$k$ model for both players until the Nash equilibrium is reached. 

Similarly, in the CH model, the reasoning step $s_k$ for $k=1,2,\ldots,$ is defined as 
\begin{equation*}
     (s_k^1,s_k^2)  = \left(p^1\cdot \sum_{j=0}^{k-1} f_k(j;\tau)s^2_j,~p^2\cdot \sum_{j=0}^{k-1} f_k(j;\tau)s^1_j\right),
\end{equation*}
where $f_k(j;\tau)$ is the conditional probability for a step-$k$ player, as defined in Equation \eqref{conditional pdf}. As before, each $s_k^i$ must be restricted to the player's choice range $[a^i,b^i]$.

The complete sets of predictions from the level-$k$ and the CH models for the different parameters (provided in Table \ref{tab:GG_parameters}) used in the GGs are given in Tables \ref{tab:GG_lk_predictions} and \ref{tab:GG_CH_predictions}, respectively. 

\medskip

\noindent\textbf{Estimation.} As with the pBCG, we estimate the (same) parameters for both the level-$k$ and CH models by MLE. The notable difference here is that we collect data sequentially for the 16 rounds of the GG. As we collect these sequence of responses 100 times for each LLM, we are able to estimate both models on an individual-response level. In other words, we estimate each model for every sequence of responses and for our final estimates, we take the average of the parameters by LLM.\footnote{Naturally, another way to approach the estimation is to aggregate the data for each round, and estimate the model for each of the 16 rounds of the GG, as if each round is their own one-shot game. However, this does not make use of the sequential nature in which the GG is played in \cite{costa-gomes_cognition_2006}, which we attempt to stay as close to as possible, in terms of experimental design.} The interpretation of both models remains the same. A detailed description of the estimation for both models is provided in Section \ref{est:gg} of Appendix B.

\subsection{Design}

Different parameters in $\left(a^i,b^i,p^i\right)$ for players $i=1,~2$ constitute different games. We follow precisely the 16 GGs used in \cite{costa-gomes_cognition_2006}. Each player has a target multiplier from $\{0.5, 0.7, 1.3, 1.5\}$, a lower limit from $\{100,300\}$, and an upper limit from $\{500,900\}$. The complete set of parameters, with the order of the games played, is provided in Table \ref{tab:GG_parameters}. For each LLM, we collect 100 sets of responses for the 16 rounds of the sequentially played GG, that is, using the same experimental design as \cite{costa-gomes_cognition_2006}. 

\medskip

\subsection{Results}
As with the pBCG, we conducted a series of preliminary questions to test whether LLMs understand the rules of the games, as well as choosing accurate best responses when given the action of the other player. All LLMs answered questions correctly with respect to rules, however only the reasoning LLMs and \claudethree consistently provided correct answers to the best-responses questions, unlike the other LLMs with approximately 100\% error rate. Given this, we collect data focusing only on the reasoning LLMs and \claudethreens.

\begin{figure}[H]
    \centering
    \includegraphics[width=0.8\textwidth]{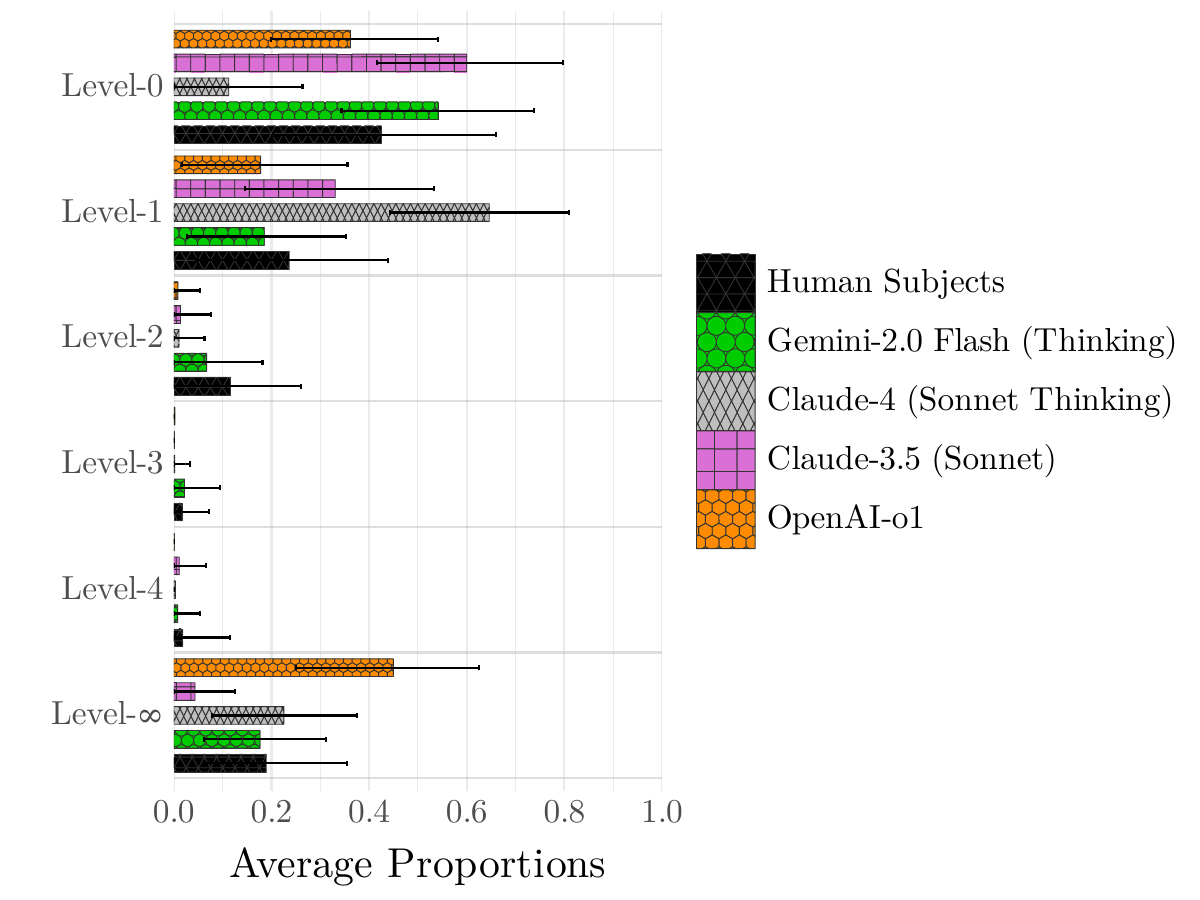}
    \caption{Estimates of the proportions of level-$k$ types for the GG with bootstrapped 95\% confidence intervals.}
    \label{fig:gg_lk}
\end{figure}

From Figure \ref{fig:gg_lk}, the majority of responses for \claudethree are consistent with $L_0$ and $L_1$ behavior with virtually no higher-order strategic reasoning. Only the reasoning LLMs exhibit any substantial proportion of $L_\infty$ behavior, with far fewer $L_0$ and $L_1$ responses for \gptoonens. The level-$k$ proportions are also estimated using data from \cite{costa-gomes_cognition_2006} ($N=88$). Estimates suggests that \geminitwo performs similarly to human subjects, with \claudethree performing worse and \gptoone and \claudefour demonstrating more strategic reasoning than human subjects, strongly in favor of \gptoonens. Table \ref{tab:gg_tests} reports frequencies of round-by-round comparisons with human subjects, showing that \gptoone and \claudefour demonstrate more strategic reasoning than \claudethree and \geminitwo in 12 and 10 out of 16 rounds, respectively. The results from the CH model are in line with the level-$k$ model. On average, \gptoone exhibits approximately 3 steps of reasoning, which is roughly 2 more steps of reasoning than \claudethree and 1 more step than \geminitwo and \claudefourns. Human subjects from \cite{costa-gomes_cognition_2006} exhibit just under 2 steps suggesting human subjects exhibit lower strategic reasoning than \gptoonens, and with comparable levels to the other reasoning LLMs.

\begin{table}[ht!]
\centering
\begin{tabular}{l|>{\centering\arraybackslash}m{2.5cm}}
\toprule
 & $\tau$ \\ \midrule
\multirow{2}{*}\gptoone   & 2.84 \\ 
           & (1.78,4.08) \\ \midrule
\multirow{2}{*}\claudethree & 0.69 \\ 
           & (0.00,1.43) \\ \midrule
\multirow{2}{*}\claudefour & 1.94 \\ 
           & (1.29,2.91) \\ \midrule
\multirow{2}{*}\geminitwo & 1.51 \\ 
           & (0.68,2.51) \\ \midrule
\multirow{2}{*}{Human Subjects} & 1.78 \\ 
           & (0.81,3.00) \\ \bottomrule
\end{tabular}
\caption{Estimates of $\tau$ from the CH model GG.}
\label{tab:baseline_CH_GG}
\end{table}

\section{Money Request Game}
The MRG offers several advantages over the pBCG and GG. From a theoretical standpoint, its simplicity triggers behavior associated with the level-\textit{k} reasoning with an intuitive $L_0$ assumption and clearly separates reasoning levels within the action space. Additionally, there is no pure Nash equilibrium, requiring players to depend on more sophisticated strategic reasoning.\footnote{Choosing any particular number with certainty for each player is not a Nash equilibrium. Instead, each player would need to randomly choose a set of strategies according to a pre-specified distribution. We do not consider such a mixed-strategy equilibrium as we only request a single action from each LLM subject, not a distribution over actions.}$^,$\footnote{We focus on Game 1 of \cite{arad_1120_2012}. Game 3, which is a variation of Game 1 is included in Section \ref{sec:appendix_mrg} of Appendix B; the results are approximately the same across games.} 

\medskip

\noindent\textbf{Game.} Two players simultaneously choose an integer between 11 and 20. Each player is awarded an amount of money corresponding to the number they choose. Additionally, if a player selects a number exactly one less than the opponent's choice, the player receives an extra payment of 20.

\medskip

\noindent\textbf{Theoretical Predictions.}
The choice of 20 plays a role as an anchor to which we denote as $L_0$. Clearly, the choice of 20 is not driven by uniform randomization, but it is a \textit{safe} choice since it does not involve any strategic concern in the sense that a player can secure the payoff of 20 from the choice of 20 (but without any chance of receiving the bonus payoff). As such, the choice of 19 is the best response to 20, the choice of 18 is the best response to 19,..., and the choice of 11 is the best response to 12. 

From this observation, the level-\textit{k} model predictions for the reasoning levels $l_k$ are as follows. 
\begin{equation*}
    l_0=20,~~l_1=19,~~l_2=18,~~ \ldots,~~l_9=11.
\end{equation*}
Similarly, in the CH model, the reasoning step $s_k$ for $k=1,2,\ldots,$ is defined as
\begin{equation*}
    s_k = \text{round}\left(\sum_{j=0}^{k-1}f_k(j;\tau)s_j\right)-1,
\end{equation*}
where $\text{round}(x)$ outputs the nearest integer to $x$, and $f_k(j;\tau)$ is the conditional probability for of a step-$k$ player, as defined in Equation \eqref{conditional pdf}.

\medskip

\noindent\textbf{Estimation.} As with the pBCG and GG, we use MLE to estimate the level-$k$ model and the CH model, analogously to what was done previously. We keep the same number of levels/steps as the pBCG but replace $L_\infty$ with $L_4$, as there is no pure Nash equilibrium. We still accommodate behavior which is not predicted by the level-$k$ or CH model by including a purely random type. However, as $L_0$ behavior can be very distinct from purely random behavior, we specifically distinguish between these two types.\footnote{Another reason for doing this is to allow 15 to be choice consistent with uniform randomization choice, and lower choices, to be considered as random behavior, rather than a very high stage of reasoning. This is consistent with the empirical results in \cite{arad_1120_2012} with a minority of subjects choosing lower than 15.} Given the finite nature of the actions and theoretical predictions in MRG, MLE is more appropriately specified using the responses and best guesses directly, as opposed to a noisy best guess. A detailed description of the estimation for both models is provided in Section \ref{est:mrg} of Appendix B.

\subsection{Results} 
Each LLM consistently provided correct answers, both when asked about the game rules and when responding to questions about best responses. 

\begin{figure}[ht!]
    \centering
    \includegraphics[width=0.8\textwidth]{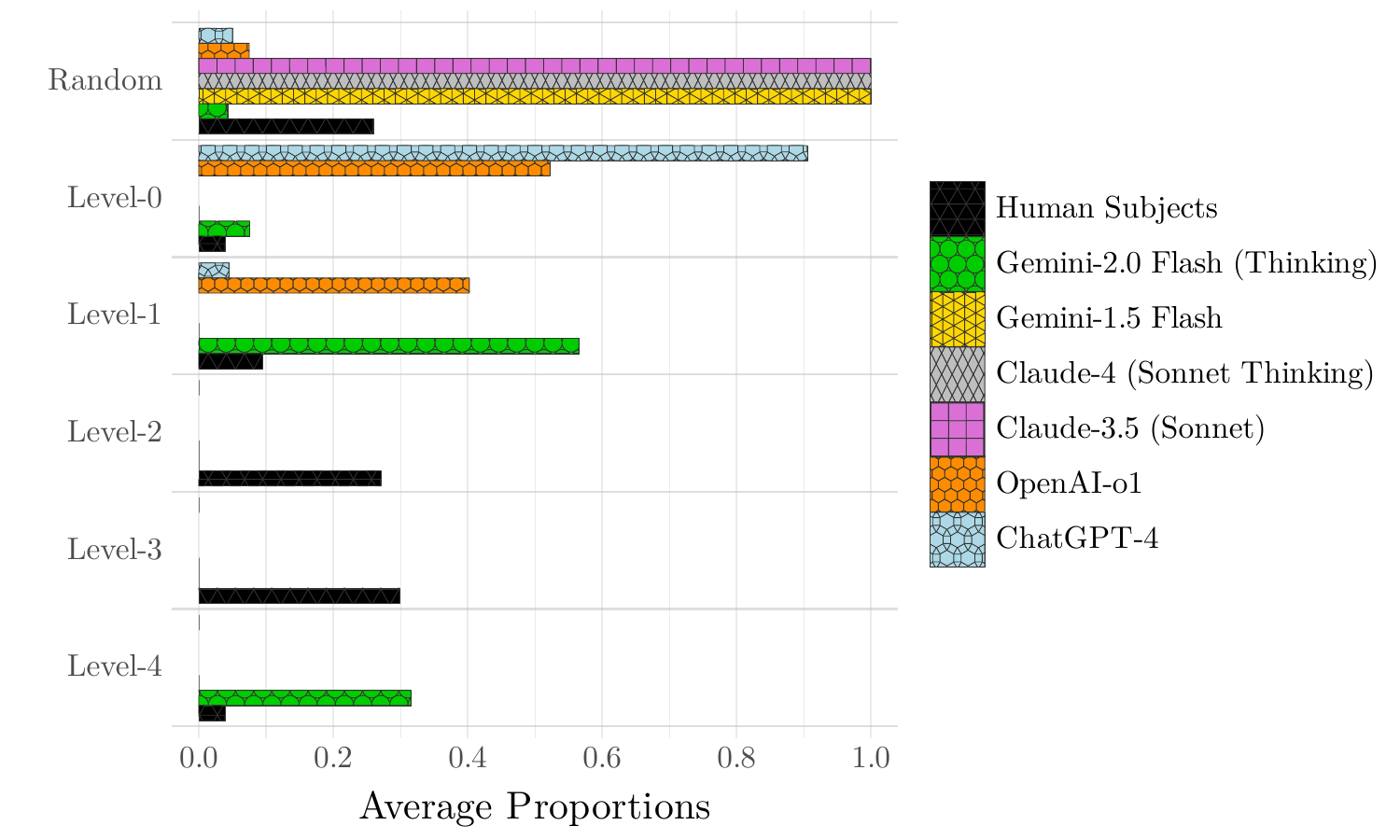}
    \caption{Estimates of the proportions of level-$k$ types for MRG.}
    \label{fig:mrg_lk}
\end{figure}

From Figure \ref{fig:mrg_lk}, the distinction between standard LLMs and reasoning LLMs is not as clearly dichotomous. While it is clear that \geminitwo exhibits the highest strategic reasoning (and overall still weakly in favor of reasoning LLMs), all other LLMs do not show any substantial strategic reasoning beyond $L_1$ with both \gptoone and \claudefour exhibit substantially high levels of random/$L_0$ behavior. Using data from \cite{arad_1120_2012} to give us the human subject benchmark suggests that they are mainly $L_2$ and $L_3$ suggesting all LLMs are vastly outperformed except \geminitwo which exhibits comparable or higher levels of reasoning than human subjects.\footnote{Using data from \cite{arad_1120_2012} ($N=108$), we also conduct similar one-sided KS tests as with the pBCG. Excluding LLMs that exhibited random behavior by our definition (\geminionens) and responses from both \claudethree and \claudefour that are random (by not exhibiting $L_0$ to $L_4$ behavior), all p-values were less than 0.005. This implies that human subject responses were cumulatively lower than LLMs, thus suggesting LLMs were outperformed across the whole distribution of responses} 

\begin{table}[]
\centering
\begin{tabular}{l|>{\centering\arraybackslash}m{2.5cm}}
\toprule
 & \text{MRG} \\ \midrule
\multirow{2}{*}\gptfour     & 0.91 \\
             & (0.86,0.97) \\ \midrule
\multirow{2}{*}\gptoone     & 1.21 \\ 
             & (1.11,1.31) \\ \midrule
\multirow{2}{*}\claudethree & 0.00 \\ 
             & (0.00,0.00) \\ \midrule
\multirow{2}{*}\claudefour   & 0.00 \\ 
             & (0.00,0.70) \\ \midrule
\multirow{2}{*}\geminione     & 0.00 \\ 
             & (0.00,0.00) \\ \midrule
\multirow{2}{*}\geminitwo     & 1.00 \\ 
             & (1.00,1.22) \\ \midrule
\multirow{2}{*}{Human Subjects}     & 3.14 \\ 
             & (2.17,3.65) \\ \bottomrule
\end{tabular}
\caption{Estimates of $\tau$ from the CH model for MRG.}
\label{tab:mrg_tau}
\end{table}

From the estimates of $\tau$ in Table \ref{tab:mrg_tau}, both \gptoone and \geminitwo exhibit the highest strategic reasoning followed by \gptfourns, with the remaining showing no steps of reasoning. These estimates are largely in agreement with the the level-$k$ model. The largest discrepancy is for \geminitwo where $\tau$ is 1, which may be surprising at first inspection. However, this is a natural consequence of the CH model itself. Recall that players are best responding to mixtures of players with lower steps of reasoning. As this mixture is governed by a (truncated) Poisson distribution, this suggests that it is empirically challenging to fit a CH model to behavior with multiple peaks in the distribution of responses. As such, the CH model with a low $\tau$ provides the best fit for this kind of data.\footnote{If there is a vast majority of answers that are consistent with higher-steps of behavior, only then would this result in a much higher estimate of $\tau$.} The low estimates of $\tau$ provide a different perspective from the level-$k$ model; the average number of steps of reasoning is still low, given that we observe a majority of lower-step responses, despite a mass of isolated high-step responses.\footnote{This might suggest that the CH model is not an appropriate model for fitting these kind of data, as we want to observe a natural hierarchy of steps of reasoning, rather than observing jumps in reasoning, ideally with one mode (peak) in the distribution of responses.} Compared to human subjects, LLMs are clearly outperformed, with human subjects exhibiting just over 3 steps of reasoning.

\section{Discussion}
In this paper, we evaluate the strategic reasoning capabilities of standard and reasoning LLMs using models of bounded rationality. In the pBCG, reasoning LLMs demonstrated substantially higher strategic sophistication than standard LLMs across most conditions with human subjects typically outperformed overall. Surprisingly, all LLMs struggled when the target multiplier exceeded one ($p=4/3$), likely due to training predominantly on games requiring downward iterative reasoning. When faced with the repeated pBCG (including when $p=4/3$), all LLMs exhibited learning and convergence toward the rational prediction, but with reasoning LLMs converging faster than standard LLMs. In the GG, a much clearer divide emerged between reasoning and standard LLMs. While reasoning LLMs exhibited substantial strategic sophistication, standard LLMs except \claudethree failed even basic comprehension checks regarding best-response calculations. Despite understanding, \claudethree was clearly outperformed by the reasoning LLMs, with reasoning LLMs exhibiting equivalent or superior strategic reasoning to human subjects. For the MRG, while reasoning LLMs still outperformed standard LLMs, human subjects exhibited superior strategic reasoning, with only \geminitwo demonstrating comparable performance to humans, while other reasoning LLMs showed surprisingly high levels of random behavior. While our analyses across the three behavioral economics games reveal that LLMs’ strategic reasoning capabilities are not uniform, our overarching conclusion is that reasoning LLMs exhibit substantially higher strategic reasoning compared to standard LLMs, with reasoning LLMs overall exhibiting superior strategic sophistication compared to typical human subjects. 

The games in our study were specifically chosen because they require players to anticipate others' strategies, allowing us to investigate, model, and evaluate LLMs' strategic reasoning \citep{zhang_k-level_2024}.\footnote{While higher-order reasoning -- even at the Nash equilibrium -- does not always lead to a better outcome in games, there is an important exception in the two-person pBCG ($n=2$): higher-order reasoning always results in a better outcome because choosing the smaller number guarantees a win in this case. Our result from the two-person pBCG aligns with the results in other pBCGs in Figure \ref{fig:lk_pb} and Table \ref{tab:ch_pb}. This observation supports our conclusion, which draws on measuring strategic reasoning using both level-$k$ and CH models.} Our results reveal a clear distinction between standard LLMs and reasoning LLMs in their ability to engage in strategic reasoning. In the cases where standard LLMs could provide accurate best responses -- demonstrating analytical ability -- when given explicit information about the strategies of the others, they struggled in multi-agent environments that require anticipating opponents’ strategies through higher-order iterated reasoning, whereas reasoning LLMs demonstrate strategic reasoning abilities that often match or exceed human performance across the games. Despite this, we still observed (surprising) drops in strategic reasoning in reasoning LLMs when given essentially logically-equivalent variations of the pBCG, which is in line with recent work by Apple \citep{shojaee_illusion_2025} showing reasoning LLMs exhibit inconsistent performance across different puzzle variations and complexity levels (in a non-strategic environment).\footnote{Recent criticisms of \cite{shojaee_illusion_2025}, such as \cite{lawsen_comment_2025},  argue that many observed failures of reasoning LLMs are a result of critical issues in experimental design (e.g. output token limits and impossible puzzles) rather than fundamental reasoning limitations. Preliminary evidence suggests that when experiments are designed that address these issues, reasoning LLMs exhibit much higher accuracy, in direct contrast with Apple's original findings. As we are not constrained by similar experimental design issues (especially as we are concerned with strategic environments), our findings are still consistent (yet distinct) from Apple.} With their growing application in AAI, reasoning LLMs have shown their potential in multi-agent reasoning, particularly in environments where anticipating others' strategies is vital \citep{duan_gtbench_2024, lore_strategic_2024, duan_reta_2024,bousetouane_agentic_2025,singh_enhancing_2024}.

An interesting finding is that even standard LLMs perform significantly better in repeated interaction than in one-shot environments. This suggests that sequential feedback enables LLMs to refine their strategic understanding through iterative learning. This insight helps explain why reasoning LLMs exhibit superior capabilities -- they are trained to effectively \textit{think}, i.e., refine strategies by breaking down tasks into more manageable, simpler steps. Through this process, reasoning LLMs gain a better understanding of any given reasoning task, analogous to the procedure in the repeated feedback of the pBCG. We believe this is the main reason for reasoning LLMs' success in strategic reasoning (particularly \gptoonens), especially compared to other LLMs that are not similarly trained.\footnote{Although \claudethree is said to exhibit higher reasoning capabilities, the nature of its reasoning is not public, therefore, we do not comment on its performance compared to \gptoone and \geminitwo which we know are designed for strategic thinking.} This paper underscores the emergence of strategic reasoning in reasoning LLMs, a natural capability that was yet to emerge in standard LLMs.

\section*{Methods}
\noindent\textbf{Collecting LLM data}

\noindent We collect data from LLMs from OpenAI, Anthropic, and Google using their APIs. The exact model names of each LLM used are: gpt-4, o1-preview-2024-09-12, claude-3-5-sonnet-20241022, claude-sonnet-4-20250514, gemini-1.5-flash, gemini-2.0-flash-thinking-exp-1219. For consistency and replicability, we used the same system message across all games, with differences only in the number of rounds played. We tell LLMs to state their final answer in [] to allow easy extraction of their final response. The rules and description of the games were provided in the assistant message. The full set of prompts is contained in Appendix B.

\noindent\textbf{Human subject data}

\noindent We use human subject data from seminal papers in behavioral economics to compare with LLMs. For the pBCG, we use data from the seminal papers, as well as other published papers for the alternate conditions \cite{nagel_unraveling_1995,hermes_if_2022,bosch-domenech_one_2002}. For the GG, we use data from the original paper \citep{costa-gomes_cognition_2006}. For the MRG, we use data from the original paper \citep{arad_1120_2012}. These serve as essential benchmarks to compare with LLM performance with their results being representative of human subject performance. As mentioned, all datasets come from published work and were subject to their own ethics committees. 

\noindent\textbf{Testing LLM and human subjects data}

\noindent Using standard MLE techniques, we estimate the level-$k$ model and the CH model. Stata estimation codes for these models can be found in \cite{moffatt_experimetrics_2015}, a textbook for estimation in behavioral and experimental economics. We translated and edited the code from Stata to R; we obtain the same results from practice datasets from \cite{moffatt_experimetrics_2015} to ensure replicability. Technical estimation details can be found in Appendix B. Further, we do distributional tests (KS tests) to compare raw responses. We use both one-sided to check for first order stochastic dominance to compare rationality and two-sided tests to compare the distribution of responses. 

\section*{Data Availability}
\noindent All datasets used are publicly available \href{https://github.com/GavinKader/LLMs-Strategic-Reasoning}{\textcolor{blue}{\textit{here}}}. This includes all estimates and figures contained in this article

\section*{Code Availability}
\noindent Our code is publicly available \href{https://github.com/GavinKader/LLMs-Strategic-Reasoning}{\textcolor{blue}{\textit{here}}}.

\bibliographystyle{aea}
\bibliography{main.bib}

\section*{Author Contributions}
\noindent G.K. and D.L wrote the paper, designed the prompts for the experiments, and interpreted the results. G.K. collected and analyzed the data, and interpreted results. G.K. and D.L share first authorship. 

\section*{Competing Interests}
\noindent The authors declare no competing interests.

\section*{Additional Information}
\noindent \textbf{Extended data} is available for this paper in Appendix A.

\noindent \textbf{Supplementary information} is available for this paper in Appendix B.

\noindent \textbf{Correspondence} should be addressed to Dongwoo Lee.

\newpage\appendix

\setcounter{page}{1}
\setcounter{table}{0}
\renewcommand{\thetable}{A\arabic{table}}

\section{Supplementary Tables}

\begin{table}[h!]
\centering
\resizebox{\columnwidth}{!}{%
\begin{tabular}{@{}l|c|c|c|c|c@{}}
\toprule
LLM                 & Game (pBCG)              & H0: Two-sided & H0: Not Less & H0: Not Greater & More rational \\ \midrule
\gptfour           & Median        & 0             & 0            & 0.243           & Human         \\
\gptfour           & n=unspecified & 0             & 0            & 0.57            & Human         \\
\gptfour           & n=2 & 0             & 0            & 0            & -         \\
\gptfour           & p=1/2         & 0             & 0            & 0.009           &  -             \\
\gptfour           & p=2/3         & 0             & 0.192        & 0               & LLM           \\
\gptfour           & p=4/3         & 0             & 0.361        & 0               & Human         \\ \midrule
\gptoone            & Median        & 0             & 1            & 0               & LLM           \\
\gptoone            & n=unspecified & 0             & 1            & 0               & LLM           \\
\gptoone            & n=2 & 0             & 1            & 0               & LLM           \\
\gptoone            & p=1/2         & 0.011         & 0.005        & 0.048           &  -        \\
\gptoone            & p=2/3         & 0             & 0.123        & 0               & LLM           \\
\gptoone            & p=4/3         & 0             & 1            & 0               & Human         \\ \midrule
\claudethree        & Median        & 0.001             & 0.001            & 0.007           &  -             \\
\claudethree        & n=unspecified & 0             & 0            & 0.010            &  -             \\
\claudethree        & n=2 & 0             & 0            & 0            &  -             \\
\claudethree        & p=1/2         & 0             & 0            & 0           &    -      \\
\claudethree        & p=2/3         & 0             & 0.028            & 0           &  -             \\
\claudethree        & p=4/3         & 0             & 0.285            & 0               &  Human             \\ \midrule
\claudefour          & Median        & 0             & 0            & 0.003           &    -      \\
\claudefour          & n=unspecified & 0             & 0            & 0           &    -      \\
\claudefour          & n=2 & 0             & 0.001            & 0           &    -      \\
\claudefour          & p=1/2         & 0             & 0.762            & 0           & LLM         \\
\claudefour          & p=2/3         & 0             & 1            & 0               &  LLM             \\
\claudefour          & p=4/3         & 0.053             & 0.027        & 0.474          & LLM         \\ \midrule
\geminione          & Median        & 0             & 0            & 0.2           & Human         \\
\geminione          & n=unspecified & 0             & 0            & 0.153           & Human         \\
\geminione          & n=2 & 0             & 0            & 0           & -         \\
\geminione          & p=1/2         & 0             & 0            & 0.555           & Human         \\
\geminione          & p=2/3         & 0             & 0            & 0           &      -         \\
\geminione          & p=4/3         & 0             & 0.113        & 0               &  Human             \\ \midrule
\geminitwo          & Median        & 0             & 0            & 0           &     -     \\
\geminitwo          & n=unspecified & 0             & 0            & 0           &     -     \\
\geminitwo          & n=2 & 0             & 0            & 0           &     -     \\
\geminitwo          & p=1/2         & 0             & 0.078            & 0           & LLM         \\
\geminitwo          & p=2/3         & 0             & 0.26            & 0           &     LLM          \\
\geminitwo          & p=4/3         & 0             & 0.522        & 0               &     Human          \\ \bottomrule
\end{tabular}%
}
\caption{KS tests for raw responses from the pBCG. The distribution of responses from human subjects is more rational if it is unambiguously first-order stochastically dominating from the 1-sided KS tests.}
\label{tab:pbcg_tests}
\end{table}

\begin{table}[h!]
\centering
\begin{tabular}{@{}l|c|c@{}}
\toprule
LLM & LLMs more rational & Rounds \\ \midrule
\gptoone & 12 & 1-11, 14 \\
\claudethree & 3 & 2, 5, 9 \\
\claudefour & 10 & 1-2, 4, 6-10, 13-14, 16 \\
\geminitwo & 4 & 1-3, 8 \\ \bottomrule
\end{tabular}
\caption{Frequency of LLMs providing more Nash equilibrium responses than human subjects (out of 16), and in which specific rounds. LLMs are considered more rational if the number of Nash equilibrium responses is strictly larger than human subjects.}
\label{tab:gg_tests}
\end{table}

\begin{table}[h!]
\centering
\begin{tabular}{@{}l|c|c|c|c@{}}
\toprule
LLM                 & H0: Two-sided & H0: Not Less & H0: Not Greater & More rational \\ \midrule
\gptfour           & 0             & 0            & 0.87            & Human         \\
\gptoone          & 0             & 0            & 1               & Human         \\
\claudethree & 0             & 0            & 0.590           &      -        \\
\claudefour & 0             & 0.833            & 0           &     -         \\
\geminione  & 0             & 0.644        & 0               &    -          \\
\geminitwo  & 0             & 0            & 0.026           & -             \\ \bottomrule
\end{tabular}%

\caption{KS tests for raw responses from the MRG. The distribution of responses from human subjects is more rational if it is unambiguously first-order stochastically dominated from the 1-sided KS tests (excluding \geminionens, \claudethreens, and \claudefourns).}
\label{tab:mrg_tests}
\end{table}

\begin{table}[h!]
\centering
\resizebox{\textwidth}{!}{%
\begin{tabular}{c|>{\centering\arraybackslash}m{2cm}|>{\centering\arraybackslash}m{2cm}|>{\centering\arraybackslash}m{2cm}|>{\centering\arraybackslash}m{2cm}|>{\centering\arraybackslash}m{2cm}|>{\centering\arraybackslash}m{2cm}>{\centering\arraybackslash}m{2cm}}
\cmidrule[\heavyrulewidth]{2-7}
 & \multicolumn{3}{c|}{Player 1} & \multicolumn{3}{c}{Player 2} \\ \cmidrule{1-7}
Games & Targets & Lower Limits & Upper Limits & Targets & Lower Limits & Upper Limits \\ \midrule
 1  & 1.3 & 300 & 900 & 1.5 & 300 & 500 \\
 2 & 0.7 & 100 & 500 & 1.5 & 100 & 500 \\
 3 & 1.5 & 100 & 900 & 0.7 & 300 & 500 \\
 4 & 1.5 & 100 & 500 & 0.5 & 100 & 900 \\ 
 5  & 0.5 & 100 & 900 & 1.5 & 100 & 500 \\
 6  & 0.5 & 100 & 900 & 0.7 & 100 & 500 \\
 7 & 1.3 & 100 & 900 & 0.7 & 300 & 900 \\
 8  & 0.5 & 100 & 900 & 0.7 & 300 & 500 \\ 
 9 & 1.5 & 100 & 500 & 0.7 & 100 & 500 \\
 10 & 0.7 & 300 & 900 & 1.3 & 100 & 900 \\
 11  & 0.7 & 300 & 500 & 0.5 & 100 & 900 \\
 12 & 0.7 & 300 & 500 & 1.5 & 100 & 900 \\ 
 13  & 1.3 & 300 & 900 & 1.3 & 300 & 900 \\
 14  & 0.7 & 100 & 500 & 0.5 & 100 & 900 \\
 15  & 1.3 & 300 & 900 & 1.3 & 300 & 900 \\
 16  & 1.5 & 300 & 500 & 1.3 & 300 & 900 \\ \bottomrule
\end{tabular}
}
\caption{Task parameters from \cite{costa-gomes_cognition_2006} used for the GG.}
\label{tab:GG_parameters}
\end{table}

\afterpage{
    \begin{landscape}
    \begin{table}[p]
    \centering
    \begin{tabular}{c|>{\centering\arraybackslash}m{1.25cm}|>{\centering\arraybackslash}m{1.25cm}|>{\centering\arraybackslash}m{1.25cm}|>{\centering\arraybackslash}m{1.25cm}|>{\centering\arraybackslash}m{1.25cm}|>{\centering\arraybackslash}m{1.25cm}|>{\centering\arraybackslash}m{1.25cm}|>{\centering\arraybackslash}m{1.25cm}|>{\centering\arraybackslash}m{1.25cm}|>{\centering\arraybackslash}m{1.25cm}|>{\centering\arraybackslash}m{1.25cm}|>{\centering\arraybackslash}m{1.25cm}}
    \toprule
    Games & $l_0$ & $l_1$ & $l_2$ & $l_3$ & $l_4$ & $l_5$ & $l_6$ & $l_7$ & $l_8$ & $l_9$ & $l_{10}$ & $l_{11}$ \\
    \midrule
    1  & 600 & 520 & $650^*$ & & & & & & & & & \\
    2 & 300 & 210 & 315 & 220.5 & 330.75 & 231.53 & 347.29 & 243.10 & $350^*$ & & & \\
    3 & 500 & 600 & 525 & 630 & 551.25 & 661.50 & 578.81 & 694.58 & 607.75 & 729.30 & 638.14 & $750^*$ \\
    4 & 300 & 500 & 225 & 375 & 168.75 & 281.25 & $150^*$ & & & & & \\
    5  & 500 & 150 & 250 & 112.50 & 187.50 & $100^*$ & & & & & & \\
    6  & 500 & 150 & 175 & $100^*$ & & & & & & & & \\
    7 & 500 & 780 & 455 & 709.80 & 414.05 & 645.92 & $390^*$ & & & & & \\
    8  & 500 & 200 & 175 & $150^*$ & & & & & & & & \\
    9 & 300 & 450 & 315 & 472.5 & 330.75 & 496.13 & 347.29 & $500^*$ & & & & \\
    10 & 600 & 350 & 546 & 318.50 & 496.86 & $300^*$ & & & & & & \\
    11  & 400 & 350 & $300^*$ & & & & & & & & & \\
    12 & 400 & 350 & 420 & 367.50 & 441 & 385.88 & 463.05 & 405.17 & 486.20 & 425.43 & $500^*$ & \\
    13  & 600 & 780 & $900^*$ & & & & & & & & & \\
    14  & 300 & 350 & 105 & 122.50 & $100^*$ & & & & & & & \\
    15  & 600 & 780 & $900^*$ & & & & & & & & & \\
    16  & 400 & $500^*$ & & & & & & & & & & \\
    \bottomrule
    \end{tabular}%
    \caption{Predictions from the level-$k$ model for the 16 GGs. Asterisk refers to reaching the Nash equilibrium action.}
    \label{tab:GG_lk_predictions}
    \end{table}
    \end{landscape}
}

\newpage
\begin{table}[h!]
\centering
\begin{tabular}{c|>{\centering\arraybackslash}m{1.25cm}|>{\centering\arraybackslash}m{1.25cm}|>{\centering\arraybackslash}m{1.25cm}|>{\centering\arraybackslash}m{1.25cm}|>{\centering\arraybackslash}m{1.25cm}|>{\centering\arraybackslash}m{1.25cm}}
\toprule
Games & $s_0$ & $s_1$ & $s_2$ & $s_3$ & $s_4$ & $s_5$ \\ \midrule
1  & 600 & 520 & 598 & 614.14 & 618.96 & 620.44 \\
2 & 300 & 210 & 273 & 268.44 & 268.26 & 268.30 \\
3 & 500 & 600 & 555 & 565.24 & 568.08 & 569.00 \\
4 & 300 & 500 & 435 & 397.76 & 387.11 & 383.69 \\
5  & 500 & 150 & 210 & 212.33 & 210.52 & 209.71 \\
6  & 500 & 150 & 165 & 145.29 & 137.59 & 134.97 \\
7 & 500 & 780 & 585 & 592.10 & 591.07 & 590.46 \\
8  & 500 & 200 & 185 & 174.14 & 170.90 & 169.89 \\
9 & 300 & 450 & 369 & 381.57 & 384.40 & 385.26 \\
10 & 600 & 350 & 467.60 & 449.57 & 444.85 & 443.36 \\
11  & 400 & 350 & $300^*$ & & & \\
12 & 400 & 350 & 392 & 390.91 & 391.55 & 391.85 \\
13  & 600 & 780 & $900^*$ & & & \\
14  & 300 & 350 & 203 & 175.84 & 165.89 & 162.55 \\
15  & 600 & 780 & $900^*$ & & & \\
16  & 400 & $500^*$ & & & & \\
\bottomrule
\end{tabular}
\caption{Example predictions from the CH model for the 16 GGs at $\tau=1.5$}
\label{tab:GG_CH_predictions}
\end{table}

\newpage
\setcounter{figure}{0}
\renewcommand{\thefigure}{B\arabic{figure}}
\section{Supplementary Methods and Notes}
\subsection{Supplementary MRG}\label{sec:appendix_mrg}
Here, we analyze Game 3 of \cite{arad_1120_2012}, which is a costless iteration version of the MRG in the main text. Game 1 (main text) and Game 2 of \cite{arad_1120_2012} are game theoretically identical. The only difference between Games 1 and 2 is the increased salience of $L_0$ behavior.\footnote{Additionally, some LLMs answered the understanding questions poorly for Game 1 when asked about best responses to 11 (the lowest number). To remedy this, we make the instructions clearer by being specific about the payoff when 11(/20) is chosen and 11(/20) is played by the other participant. As such, adding this to Game 1's instructions essentially replicates Game 2's instructions.} The payoff structure of Game 3 is a slight variation from Game 1/2, with a different equilibrium prediction, but is ultimately similar to MRG.

\medskip

\noindent\textbf{MRG (Game 3).} Two players simultaneously choose an integer between 11 and 20. Each player is awarded an amount of money corresponding to the number they choose. A player is awarded 20 if they choose 20. Choosing any other number results in receiving 17. Additionally, if a player selects a number exactly one less than the opponent's choice, the player receives an extra payment of 20.
\medskip

\begin{figure}[h!]
    \centering
    \includegraphics[width=0.8\textwidth]{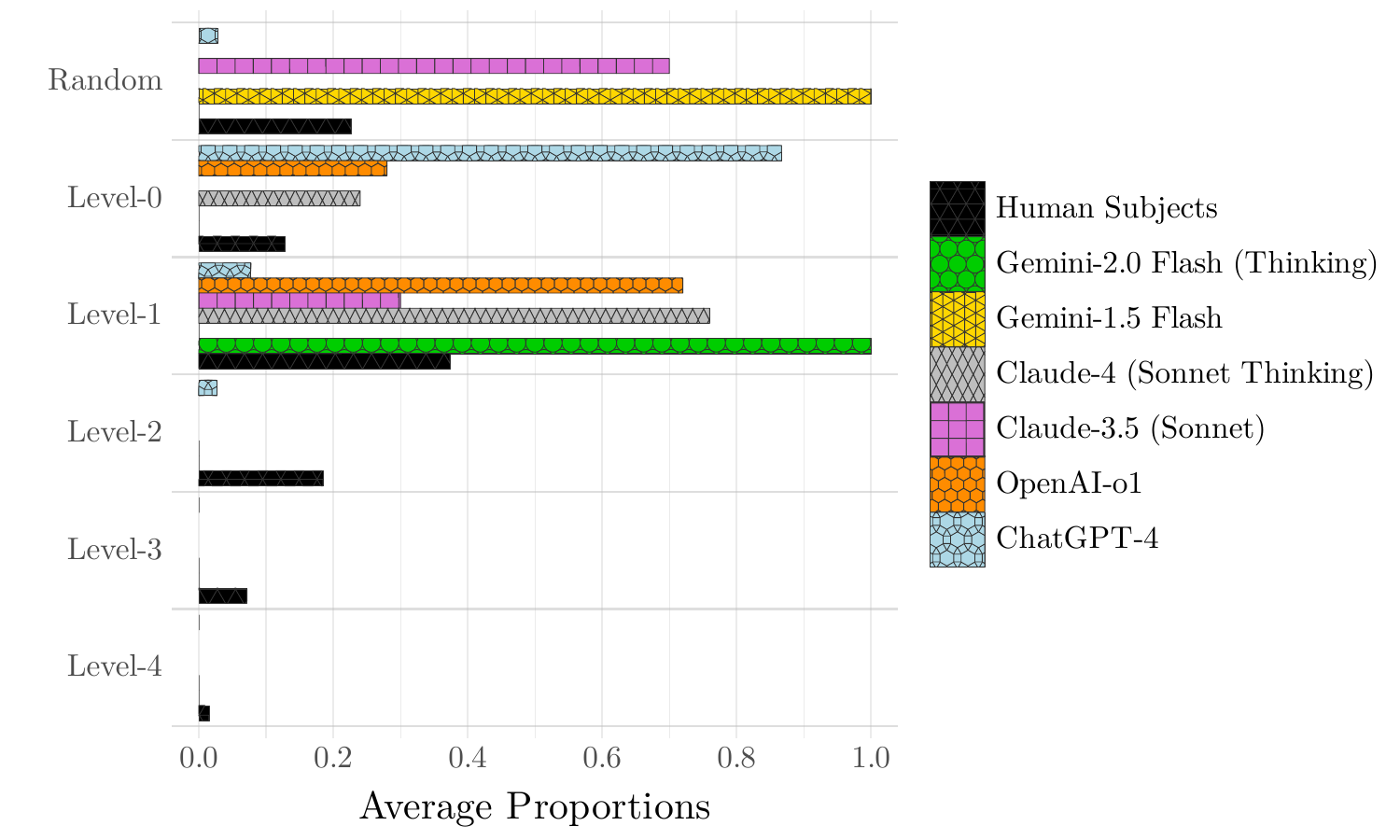}
    \caption{Estimates of the proportions of level-$k$ types for MRG (Game 3).}
    \label{fig:mrg3_lk}
\end{figure}

\begin{table}[h!]
\centering
\begin{tabular}{l|>{\centering\arraybackslash}m{2.5cm}}
\toprule
 & \text{MRG (Game 3)} \\ \midrule
\multirow{2}{*}\gptfour     & 1.00 \\ 
             & (0.93,1.06) \\ \midrule
\multirow{2}{*}\gptoone     & 1.60 \\ 
             & (1.53,1.67) \\ \midrule
\multirow{2}{*}\claudethree & 0.00 \\ 
             & (0.00,0.00) \\ \midrule
\multirow{2}{*}\claudefour   & 1.64 \\ 
             & (1.57,1.70) \\ \midrule
\multirow{2}{*}\geminione     & 0.00 \\ 
             & (0.00,0.00) \\ \midrule
\multirow{2}{*}\geminitwo     & 1.89 \\ 
             & (1.89,1.89) \\ \bottomrule
\end{tabular}
\caption{Estimates of $\tau$ from the CH model for MRG (Game 3).}
\label{tab:mrg3_tau}
\end{table}

For the MRG (Game 3), the level-$k$ results are largely the same for \gptfourns, and \geminione as with the MRG. \geminitwo is less sophisticated, with the majority of answers being consistent with $L_1$ behavior, with a large reduction in higher-step responses (Figure \ref{fig:mrg3_lk}). For \gptoonens, \claudethreens, and \claudefour we observe increases in strategic reasoning which is in line with the original formulation of MRG (Game 3). This is because undercutting as a strategy is no longer costly. Unlike in MRG when players may balance between choosing a higher number (between 11 and 19) yielding a higher payoff, and undercutting to get the bonus payoff, MRG (Game 3) removes the marginal incentive in choosing a higher number. This suggests that we should observe increases in strategic reasoning in MRG (Game 3) relative to MRG if there is sufficient understanding in the changed payoff structure, which we only observe for \gptoonens, \claudethreens, and \claudefour.\footnote{Confirmed by one-sided KS tests at 5\% significance level.} When compared with \cite{arad_1120_2012} ($N=53$), human subjects are mainly random, $L_1$ and $L_2$ suggesting LLMs are all still vastly outperformed by human subjects.\footnote{The estimated level-$k$ model proportions (random, $L_1$ to $L_4$) are $0.23, 0.13, 0.37, 0.18, 0.07, 0.02$, respectively.}$^{,}$\footnote{Conducting similar one-sided KS tests with data from (again excluding \geminione and \claudethreens) shows that human subjects outperformed LLMs (Table \ref{tab:mrg_tests_game3}).}

The estimates of $\tau$ are in line with the level-$k$ explanations above (Table \ref{tab:mrg3_tau}).\footnote{The CH model estimate from human subjects is $\tau = 2.45; [1.16,2.45]_\text{95\% CI}$.} \claudefour and \geminitwo have apparent jumps in steps of reasoning owing to the fact that there is now a clear peak in the distribution of responses around the step-1/2 responses, allowing a much better fit of the CH model, given the Poisson distribution assumption.\footnote{A similar explanation for why \claudethree does not exhibit higher steps of reasoning applies, despite showing an increase in $L_1$ responses. As the majority of the answers are consistent with random behavior, with no responses in between random and $L_1$ behavior, the CH model fits better for a low $\tau$.}

\begin{table}[h!]
\centering
\begin{tabular}{@{}l|c|c|c|c@{}}
\toprule
LLM                 & H0: Two-sided & H0: Not Less & H0: Not Greater & More rational \\ \midrule
\gptfour           & 0             & 0            & 1               & Human         \\
\gptoone          & 0             & 0            & 1               & Human         \\
\claudethree & 0.024         & 0.016        & 0.069           &      -        \\
\claudefour          & 0         & 0        & 1               & Human         \\
\geminione  & 0             & 0.619        & 0               &   -           \\
\geminitwo  & 0             & 0            & 0.009           & -             \\ \bottomrule
\end{tabular}%

\caption{KS tests for raw responses from the MRG (Game 3). The distribution of responses from human subjects is more rational if it is unambiguously first-order stochastically dominated from the 1-sided KS tests (excluding \geminione and \claudethreens).}
\label{tab:mrg_tests_game3}
\end{table}

\subsection{Understanding Questions}\label{sec:understanding}
The following lists all of the understanding questions asked to each LLMs related to best responses (BR) and the rules of each game. The system prompt is the same for all questions (as written in the paper). Each question is repeatedly asked in isolation (no history of previous answers).
\subsubsection{pBCG}

\begin{Box1}[]{BR for 2-person (p=2/3)}
Including yourself, there are 2 participants in total. The other participant you will be playing with is a human decision maker. All participants will be asked to pick a number between 0 and 100 inclusive. The winner will be the one choosing the number closest to 2/3 multiplied the average of all the numbers provided by the participants (including your own). For example, if you choose 10 and your opponent chooses 20, then you will be the winner, as 10 is closer to 15*2/3=10 (as 15 is the average and 2/3 is the target). Or, if you choose 30 and your opponent chooses 20, your opponent will be the winner, as 20 is closer to 25*2/3=16.67 (as 25 is the average and 2/3 is the target). In case of multiple winners, only one participant will be randomly chosen to win. Suppose your opponent chooses 20. What number from 0 to 100 earns you the most?
\end{Box1}

\noindent Best responses: \{0,1,…,19\}

\medskip

\begin{Box1}[]{BR for 11-person (p=1/2)}
Including yourself, there are 11 participants in total. The other participants you will
be playing with are human decision makers. All participants will be asked to pick a number between 0 and 100 inclusive. The winner will be the one choosing the number closest to 1/2 multiplied by the average of all the numbers provided by the participants (including your own). For example, if you choose 14 and your opponents choose 10, 22, 23, 30, 11, 16, 37, 18, 19, 20, then the opponent with 10 will be the winner, as 10 is closer to 20*1/2=10 (as 10 is the average and 1/2 is the target). Or, if you choose 25 and your opponents choose 5, 10, 20, 40, 50, 60, 70, 80, 90, 100, then you will be the winner, as 25 is closer to 50*1/2=25 (as 50 is the average and 1/2 is the target). In case of multiple winners, only one participant will be randomly chosen to win. Suppose your opponents choose 0, 80, 43, 70, 21, 33, 37, 18, 50, 50. What number from 0 to 100 earns you the most?
\end{Box1}

\noindent Best responses: \{19,20\}

\medskip

\begin{Box1}[]{BR for Baseline: 11-person (p=2/3)}
Including yourself, there are 11 participants in total. The other participants you will
be playing with are human decision makers. All participants will be asked to pick a number between 0 and 100 inclusive. The winner will be the one choosing the number closest to 2/3 multiplied by the average of all the numbers provided by the participants (including your own). For example, if you choose 14 and your opponents choose 10, 22, 23, 30, 11, 16, 37, 18, 19, 20, then you will be the winner, as 14 is closer to 20*2/3=13.33 (as 20 is the average and 2/3 is the target). Or, if you choose 25 and your opponents choose 5, 10, 20, 40, 50, 60, 70, 80, 90, 100, then the opponent with 40 will be the winner, as 40 is closer to 50*2/3=33.3 (as 50 is the average and 2/3 is the target). In case of multiple winners, only one participant will be randomly chosen to win. Suppose your opponents choose 0, 80, 43, 70, 21, 33, 37, 18, 50, 50. What number from 0 to 100 earns you the most?
\end{Box1}

\noindent Best responses: \{22,23,…,31\}

\medskip

\begin{Box1}[]{BR for 11-person (p=4/3)}
Including yourself, there are 11 participants in total. The other participants you will
be playing with are human decision makers. All participants will be asked to pick a number between 0 and 100 inclusive. The winner will be the one choosing the number closest to 4/3 multiplied by the average of all the numbers provided by the participants (including your own). For example, if you choose 14 and your opponents choose 10, 22, 23, 30, 11, 16, 37, 18, 19, 20, then the opponent with 30 will be the winner, as 30 is closer to 20*4/3=26.67 (as 20 is the average and 4/3 is the target). Or, if you choose 25 and your opponents choose 5, 10, 20, 40, 50, 60, 70, 80, 90, 100, then the opponent with 70 will be the winner, as 70 is closer to 50*4/3 (as 50 is the average and 4/3 is the target). In case of multiple winners, only one participant will be randomly chosen to win. Suppose your opponents choose 0, 80, 43, 70, 21, 33, 37, 18, 50, 50. What number from 0 to 100 earns you the most?
\end{Box1}

\noindent Best responses: \{51,52,…,62\}

\medskip

\begin{Box1}[]{BR for $\mathbf{n}$-person (p=2/3)}
Including yourself, there is a finite but unknown number $n$ of participants in total. The other participants you will be playing with are human decision makers. All participants will be asked to pick a number between 0 and 100 inclusive. The winner will be the one choosing the number closest to 2/3 multiplied by the average of all the numbers provided by the participants (including your own). For example, if you choose 14 and your opponents choose 10, 22, 23, 30, 11, 16, 37, 18, 19, 20, then you will be the winner, as 14 is closer to 20*2/3=13.33 (as 20 is the average and 2/3 is the target). Or, if you choose 25 and your opponents choose 5, 10, 20, 40, 50, 60, 70, 80, 90, 100, then the opponent with 40 will be the winner, as 40 is closer to 50*2/3=33.3 (as 50 is the average and 2/3 is the target). In case of multiple winners, only one participant will be randomly chosen to win. In the case with $n=11$, if your opponents choose 0, 80, 43, 70, 21, 33, 37, 18, 50, 50, what number from 0 to 100 earns you the most?
\end{Box1}

\noindent Best responses: \{22,23,…,31\}

\medskip

\begin{Box1}[]{BR for Median}
Including yourself, there are 11 participants in total. The other participant you will be playing with is a human decision maker. All participants will be asked to pick a number between 0 and 100 inclusive. The winner will be the one choosing the number closest to 2/3 multiplied by the median of all the numbers provided by the participants (including your own). For example, if you choose 80 and your opponents choose 0, 80, 43, 70, 21, 33, 37, 18, 50, 50, then one of the opponents with 50 will be the winner, as 50 is closer to 43(median)*2/3. Or, if you choose 100 and your opponents choose 5, 10, 20, 40, 50, 60, 70, 80, 90, 100, then the opponent with 40 will be the winner, as 40 is closer to 60(median)*2/3. In case of multiple winners, only one participant will be randomly chosen to win. Suppose your opponents choose 0, 80, 43, 70, 21, 33, 37, 18, 50, 50. What number from 0 to 100 earns you the most?
\end{Box1}

\noindent Best responses: \{21,23,…,28\}

\medskip

For each of the conditions above, LLMs were also asked the following questions (repeatedly, and in isolation) related to the rules and the win conditions of the pBCG:

\begin{Box1}[]{Rules for pBCG}
\begin{enumerate}[topsep=0pt,itemsep=-1ex,partopsep=1ex,parsep=1ex]
    \item What numbers can you choose between?
    \item If the number you choose is closest to $\{p\}$'s of the \{average/median\}, do you win the game?
    \item How many participants are there, including yourself?
    \item Suppose there are 3 participants, including you, and you have to choose the number closest to $\{p\}$ of the average. The other 2 participants chose 90 and 60. Suppose you chose 30. Therefore $\{p\}$'s of the \{average/median\} is $\{\{p\}\times 60\}$. Do you win or lose?\footnote{The question uses actual values for $\{p\}$ and \{average/median\} depending on the condition. $\{p\}\times 60$ is evaluated, i.e., the actual calculation is shown. }
    \item Will decisions made in previous rounds have any effect on the current round? (repeated pBCG)
\end{enumerate}
\end{Box1}

\bigskip

\subsubsection{GG}

\begin{Box1}[]{BR for GG}
This game concerns a decision situation in which you and another person we call ``s/he" (which will refer to a new person each round) separately and independently make decisions called GUESSES. Together, your and her/his guesses determine the numbers of POINTS that you and s/he earn in a round, which may be different. To choose your guesses, it may help you to understand how your and her/his guesses will determine the numbers of points that you and s/he earn in the decision situations. In each decision situation, each person has her/his own TARGET, LOWER LIMIT and UPPER LIMIT. These targets and limits may be different for you and her/him, and they may change from round to round. Otherwise, the decision situations are identical in all 16 rounds. Your and her/his targets, lower limits, and upper limits will be known to you and her/him every round. Both you and s/he will receive the same instructions and have the same information about the decision situations and the same access to your and her/his targets and limits. You (respectively, s/he) can choose your (her/his) guesses only within your (her/his) given limits for each round as explained below. After submitting guesses, you earn whichever is larger, ether 0 points or 200 points minus the distance between YOUR guess and the product of YOUR target times HER/HIS guess, PLUS whichever is larger, either 0 points or 100 points minus one-tenth (1/10th) the distance between YOUR guess and the product of YOUR target times HER/HIS guess. S/he earns whichever is larger, either 0 points or 200 points minus the distance between (1)HER/HIS guess and (2)the product of HER/HIS target times YOUR guess, PLUS whichever is larger, either 0 points or 100 points minus one-tenth (1/10th) the distance between (3)HER/HIS guess] and (4)the product of HER/HIS target times YOUR guess. That is, \[\text{Points} = \max\{0,200-\text{distance}\}+\max\{0,100-0.1*\text{distance}\},\]
where distance is the absolute difference between Your guess and your Target*(His/Her Guess)

This way of determining the number of points that you and s/he earn makes the number you earn larger, the closer your guess is to your target times her/his guess; and it makes the number s/he earns larger, the closer her/his guess is to her/his target times your guess. Only the distance matters, NOT whether the difference is positive or negative. You earn the same number of points when your guess is too high by a given amount as when it is too low by the same amount. It is important to understand how your (respectively, her/his) original guesses should be chosen to stay within your (her/his) limits. This will be done as follows. If your original guess is below your lower limit, then your guess should be adjusted UP to your LOWER limit; and if your original guess is above your upper limit, then your guess should be adjusted DOWN to your UPPER limit. If, for example, your lower limit is 400 and you original guess 300, then your adjusted guess is 400. If your upper limit is original 600 and your guess 900, then your adjusted guess is 600. Her/his guesses are adjusted up or down to her/his lower or upper limits in the same way, except that her/his limits may be different. Remember you and s/he can choose any number within your and her/his limits, respectively. 

\noindent Suppose that:

\noindent Your Limits \& Target are: Lower Limit = 200, Upper Limit = 600, Target = 1.2\\
\noindent Her/His Limits \& Target are: Lower Limit = 400, Upper Limit = 800, Target = 0.8

\begin{enumerate}[topsep=0pt,itemsep=-1ex,partopsep=1ex,parsep=1ex]
\item If s/he guesses 500, which of your guesses earns you the most points? How many points would you earn by entering that guess?
\item If you guess 400, which of her/his guesses earns her/him the most points? How many points would s/he earn by entering that guess?
\item If s/he guesses 800, which of your guesses earns you the most points?
\item If your guess is 600, which of her/his guesses earns her/him the most points?
\end{enumerate}
\end{Box1}
\noindent The following are the correct responses:
\begin{enumerate}[topsep=0pt,itemsep=-1ex,partopsep=1ex,parsep=1ex]
    \item Answer = 600, Points = 300;
    \item Answer = 400, Points = 212;
    \item Answer = 600;
    \item Answer = 480.
\end{enumerate}

\noindent Each of these 4 questions are repeatedly asked in isolation with the same set of instructions above (no history of previous answers for any of the questions).

LLMs were also asked the following questions (repeatedly, and in isolation) related to the rules and the win conditions of the pBCG:

\begin{Box1}[]{Rules for GG}
\begin{enumerate}[topsep=0pt,itemsep=-1ex,partopsep=1ex,parsep=1ex]
    \item What numbers can you choose between?
    \item Do you earn more/fewer/the same points when your guess is too high or too low by the same amount?
    \item Will the targets, lower limits, and upper limits for each player be known to all players?
    \item How many participants are there, including yourself?
    \item Will decisions made in previous rounds have any effect on the current round? 
\end{enumerate}
\end{Box1}

\bigskip

\subsubsection{MRG}

\begin{Box1}[]{BR for MRG}
You are randomly matched to play a game against a human decision maker. In the game, each of you chooses an integer between 11 and 20. Each participant will receive points in the amount they have chosen. A participant will receive an additional 20 points if they choose a number exactly one less chosen by the other player. For example, if the other participant chooses X, you will get (X-1)+20 points if you choose X-1 points. For example, if your opponent chooses 14, you will get 33 points if you choose 13. Or, if the other participant chooses 11, you will get 20 points if you choose 20.

\begin{enumerate}[topsep=0pt,itemsep=-1ex,partopsep=1ex,parsep=1ex]
    \item Suppose your opponent chooses 15. What number from 11 to 20 earns you the most?
    \item Suppose your opponent chooses 11. What number from 11 to 20 earns you the most?
\end{enumerate}
\end{Box1}
\noindent The following are the correct responses:
\begin{enumerate}[topsep=0pt,itemsep=-1ex,partopsep=1ex,parsep=1ex]
    \item Answer= 14;
    \item Answer = 20.
\end{enumerate}

\medskip

\begin{Box1}[]{BR for MRG (Game 3)}
You are randomly matched to play a game against a human decision maker. In the game, each of you chooses an integer between 11 and 20. If a participant chooses the number 20, they will receive 20 points (regardless of the other player’s choice). Otherwise, if a participant chooses any other number in this range, they will receive three points less than in the case where they choose 20. However, they will receive an additional amount of 20 points if they choose a number that is lower by exactly one than the number chosen by the other player. For example, if the other participant chooses X, you will get 17+20 points if you choose X-1. For example, if your opponent chooses 14, you will get 17+20 points if you choose 13. Or, if the other participant chooses 11 points, you will get 20 points if you choose 20. 

\begin{enumerate}[topsep=0pt,itemsep=-1ex,partopsep=1ex,parsep=1ex]
    \item Suppose your opponent chooses 15. What number from 11 to 20 earns you the most?
    \item Suppose your opponent chooses 11. What number from 11 to 20 earns you the most?
\end{enumerate}
\end{Box1}
\noindent The following are the correct responses:
\begin{enumerate}[topsep=0pt,itemsep=-1ex,partopsep=1ex,parsep=1ex]
    \item Answer= 14;
    \item Answer = 20.
\end{enumerate}
\noindent Every question, for both games, is repeatedly asked in isolation with the same set of instructions above (no history of previous answers, for any of the questions).

For each of the games above, LLMs were also asked the following questions (repeatedly, and in isolation) related to the rules and the win conditions of each game:

\begin{Box1}[]{Rules for MRG}
\begin{enumerate}[topsep=0pt,itemsep=-1ex,partopsep=1ex,parsep=1ex]
    \item What numbers can you choose between?
    \item How many participants are there, including yourself?
    \item Suppose you choose 19 and the participant you have been matched with chooses 20. How many points will you receive?
\end{enumerate}
\end{Box1}

\subsection{Prompts}\label{sec:tasks}

To collect data, we use the respective application programming interfaces (API) with Python/R from OpenAI, Anthropic, and Google. The exact model names of each LLM used are: gpt-4, o1-preview-2024-09-12, claude-3-5-sonnet-20241022, claude-sonnet-4-20250514, gemini-1.5-flash, gemini-2.0-flash-thinking-exp-1219. Each input prompt includes the specification of a combination of three roles: a system role, an assistant, or a user. The system role is used to guide the behavior of the LLM by providing high-level instructions and/or context. The assistant refers to the LLM itself which generates responses for each input prompt. The user role represents the queries or prompts that are used to interact with the LLM (i.e., with the assistant). By repeated specifications of the assistant and user role within each input prompt, we are able to, in effect, have a contextual conversation.  For our purpose, each input prompt will include a system role, which will be used to specify global instructions and context, and a user role, which will specify the specific rules and the request for an answer for the pBCG.\footnote{Once interacted with the API using an input prompt, the answer provided by the LLM is always specified by the assistant role.}  

Once we knew LLMs had as good an understanding of the games as was naturally and feasibly possible, given the above questions (particularly the best response questions), we kept the wording of the prompts as close to the understanding questions. For completion, we include all prompts below.
\subsubsection{pBCG}
As this is a completely standard game in the behavioral economics, instructions are based closely on instructions from the literature.\footnote{See \cite{nagel_unraveling_1995,ho_iterated_1998,bosch-domenech_one_2002,weber_learning_2003} and numerous other papers for their accompanying instructions.} We avoid monetary payoffs as rewards to avoid issues with incentives. We also provide examples in the instructions as LLMs had better answers to the best responses understand questions (as this helps LLMs understand what kind of calculations are involved).

\subsubsection{1-shot pBCG}
For the 1-shot pBCGs, the system prompt is 
\begin{Box1}[]{System}
I want you to act as a human decision maker, You will be participating in 1 round of a game. You should use your best judgment to come up with responses that you like most. Always write your final answer in [ ].
\end{Box1}
\noindent This is identical for all 1-shot pBCG. 

The following prompts include all the details of the pBCG for each condition, specified by the user role.
\begin{Box1}[]{User for 2-person (p=2/3)}
Including yourself, there are 2 participants in total. The other participant you will be playing with is a human decision maker. All participants will be asked to pick a number between 0 and 100 inclusive. The winner will be the one choosing the number closest to 2/3's of the average of all the numbers provided by the participants (including your own). For example, if you choose 10 and your opponent chooses 20, then you will be the winner, as 10 is closer to 15(average)*2/3. Or, if you choose 30 and your opponent chooses 20, your opponent will be the winner, as 20 is closer to 25(average)*2/3. In case of multiple winners, only one participant will be randomly chosen to win. Now let's start this game. Please pick one number between 0 and 100 inclusive.
\end{Box1}

\medskip

\begin{Box1}[]{User for 11-person (p=1/2)}
Including yourself, there are 11 participants in total. The other participant you will be playing with are human decision makers. All participants will be asked to pick a number between 0 and 100 inclusive. The winner will be the one choosing the number closest to 1/2 of the average of all the numbers provided by the participants (including your own). For example, if you choose 14 and your opponents choose 10, 22, 23, 30, 11, 16, 37, 18, 19, 20, then the opponent with 10 will be the winner, as 10 is closer to 20(average)*1/2. Or, if you choose 25 and your opponents choose 90, 100, 50, 20, 60, 80, 5, 70, 10, 40, then you will be the winner, as 25 is closer to 50(average)*1/2. In case of multiple winners, only one participant will be randomly chosen to win. Now let's start this game. Please pick one number between 0 and 100 inclusive.
\end{Box1}

\medskip

\begin{Box1}[]{User for Baseline: 11-person (p=2/3)}
Including yourself, there are 11 participants in total. The other participants you will be playing with are a human decision makers. All participants will be asked to pick a number between 0 and 100 inclusive. The winner will be the one choosing the number closest to 2/3's of the average of all the numbers provided by the participants (including your own). For example, if you choose 14 and your opponents choose 10, 22, 23, 30, 11, 16, 37, 18, 19, 20, then you will be the winner, as 14 is closer to 20(average)*2/3. Or, if you choose 25 and your opponents choose 90, 100, 50, 20, 60, 80, 5, 70, 10, 40, then the opponent with 40 will be the winner, as 40 is closer to 50(average)*2/3. In case of multiple winners, only one participant will be randomly chosen to win. Now let's start this game. Please pick one number between 0 and 100 inclusive.
\end{Box1}

\medskip

\begin{Box1}[]{User for 11-person (p=4/3)}
Including yourself, there are 11 participants in total. The other participants you will be playing with are a human decision makers. All participants will be asked to pick a number between 0 and 100 inclusive. The winner will be the one choosing the number closest to 4/3's of the average of all the numbers provided by the participants (including your own). For example, if you choose 14 and your opponents choose 10, 22, 23, 30, 11, 16, 37, 18, 19, 20, then the opponent with 30 will be the winner, as 30 is closer to 20(average)*4/3. Or, if you choose 25 and your opponents choose 90, 100, 50, 20, 60, 80, 5, 70, 10, 40, then the opponent with 70 will be the winner, as 70 is closer to 50(average)*4/3. In case of multiple winners, only one participant will be randomly chosen to win. Now let's start this game. Please pick one number between 0 and 100 inclusive.
\end{Box1}

\medskip

\begin{Box1}[]{User for $\mathbf{n}$-person (p=2/3)}
The other participants you will be playing with are human decision makers. All participants will be asked to pick a number between 0 and 100 inclusive. The winner will be the one choosing the number closest to 2/3's of the average of all the numbers provided by the participants (including your own). For example, suppose there are 11 participants (including yourself). In this situation, if you choose 14 and your opponents choose 10, 22, 23, 30, 11, 16, 37, 18, 19, 20, then you will be the winner, as 14 is closer to 20(average)*2/3. Or, if you choose 25 and your opponents choose 5, 10, 20, 40, 50, 60, 70, 80, 90, 100, then the opponent with 40 will be the winner, as 40 is closer to 50(average)*2/3. In case of multiple winners, only one participant will be randomly chosen to win. Now let's start this game. Please pick one number between 0 and 100 inclusive.
\end{Box1}

\medskip

\begin{Box1}[]{User for Median}
Including yourself, there are 11 participants in total. The other participants you will be playing with are a human decision makers. All participants will be asked to pick a number between 0 and 100 inclusive. The winner will be the one choosing the number closest to 2/3's of the median of all the numbers provided by the participants (including your own). For example, if you choose 14 and your opponents choose 10, 22, 23, 30, 11, 16, 37, 18, 19, 20, then the opponent with 11 will be the winner, as 11 is closer to 20(median)*2/3. Or, if you choose 12 and your opponents choose 90, 100, 50, 20, 60, 80, 5, 70, 10, 40, then the opponent with 40 will be the winner, as 40 is closer to 50(median)*2/3. In case of multiple winners, only one participant will be randomly chosen to win. Now let's start this game. Please pick one number between 0 and 100 inclusive.
\end{Box1}

\bigskip

\subsubsection{Repeated pBCG}
For the repeated pBCG in which each of the LLMs play within a group of 10 other LLM subjects of the same model, the system prompt is 
\begin{Box1}{System}
I want you to act as a human decision maker, You will be participating in 10 rounds of a game. All rounds are independent of each other meaning decisions made in each round do not affect subsequent rounds. You should use your best judgment to come up with responses that you like most. Always write your final answer in [ ].
\end{Box1}

For the first round, the user prompt is 
\begin{Box1}{User}
Including yourself, there are 11 participants in total. The other participants you will be playing with are human decision makers. For each round, you will play with the same set of participants. All participants will be asked to pick a number between 0 and 100 inclusive. The winner will be the one choosing the number closest to 4/3's of the average of all the numbers provided by the participants (including your own). For example, if you choose 14 and your opponents choose 10, 22, 23, 30, 11, 16, 37, 18, 19, 20, then the opponent with 30 will be the winner, as 30 is closer to 20(average)*4/3. Or, if you choose 25 and your opponents choose 90, 100, 50, 20, 60, 80, 5, 70, 10, 40, then the opponent with 70 will be the winner, as 70 is closer to 50(average)*4/3. In case of multiple winners, only one participant will be randomly chosen to win. Now let's start this game. After each round, all participants will be told the average, and the target (4/3*average). You will also be privately informed of whether you won or lost at the end of each round. Round 1. Please pick one number between 0 and 100 inclusive.\footnote{The same structure for the prompt is used for p=2/3, replacing the examples with the p=2/3 examples used as in the 1-shot pBCG case.}
\end{Box1}

We keep the whole history for each LLM subject in subsequent rounds by adding to the vector of messages (i.e., stacking messages) as the input to the API. After the user prompt from the first round, the answer given by the LLM is added to the messages in the assistant role.\footnote{It is trivial to extract the actual response given the system prompt requests that the final answer is given in square brackets [].} After this, as a user role message, we prompt with 

\begin{Box1}[]{User}
In the previous round, the average was *average* and the target was *average*$\times$p. You *won /lost* in the previous round.
\end{Box1}
\noindent where *average*, target, and winning (or not) are calculated based on the LLM responses given in the previous round. We then add another user role message with the prompt 
\begin{Box1}[]{User}
Round 2. Please pick one number between 0 and 100 inclusive.
\end{Box1}
\noindent We repeat this process of messages, keeping the entire history of messages for each LLM subject until round 10. Note that the history given to each LLM subject is specific to that LLM subject; no other information is shared with other LLM subjects beyond the average and the target. In this way, the system message, and the rules of the game are always contained in the first 2 messages of the history, as well as the entire history of choices that each LLM subject makes, with accompanying feedback, specific to each LLM subject. 

\bigskip

\subsubsection{GG}
The instructions come directly from \cite{costa-gomes_cognition_2006}, with minimal modifications. The system prompt is 
\begin{Box1}[]{System}
I want you to act as a human decision maker, You will be participating in 16 rounds of a game. All rounds are independent of each other meaning decisions made in each round do not affect subsequent rounds. You should use your best judgment to come up with responses that you like most. Always write your final answer in [ ].
\end{Box1}

\medskip

\noindent\textbf{GG. } For the first round, the user prompt is 
\begin{Box1}{User}
In each round, you will be matched with one of the other participants, a new one in each round. You will not know which of the other participants you are matched with, and the other participants are human decision makers. Each round concerns a decision situation in which you and another person we call `s/he' (which will refer to a new person each round) separately and independently make decisions called GUESSES. Together, your and her/his guesses determine the numbers of POINTS that you and s/he earn in a round, which may be different. To choose your guesses, it may help you to understand how your and her/his guesses will determine the numbers of points that you and s/he earn in the decision situations. In each decision situation, each person has her/his own TARGET, LOWER LIMIT and UPPER LIMIT. These targets and limits may be different for you and her/him, and they may change from round to round. Otherwise, the decision situations are identical in all 16 rounds. Your and her/his targets, lower limits, and upper limits will be known to you and her/him every round. Both you and s/he will receive the same instructions and have the same information about the decision situations and the same access to your and her/his targets and limits. You (respectively, s/he) can choose your (her/his) guesses only within your (her/his) given limits for each round as explained below. After submitting guesses, you earn whichever is larger, ether 0 points or 200 points minus the distance between YOUR guess and the product of YOUR target times HER/HIS guess, PLUS whichever is larger, either 0 points or 100 points minus one-tenth (1/10th) the distance between YOUR guess and the product of YOUR target times HER/HIS guess. S/he earns whichever is larger, either 0 points or 200 points minus the distance between (1)HER/HIS guess and (2)the product of HER/HIS target times YOUR guess, PLUS whichever is larger, either 0 points or 100 points minus one-tenth (1/10th) the distance between (3)HER/HIS guess and (4)the product of HER/HIS target times YOUR guess. That is, Points = max{0,200-distance}+max{0,100-0.1*distance},where distance is the absolute difference between Your guess and your Target*(His/Her Guess). This way of determining the number of points that you and s/he earn makes the number you earn larger, the closer your guess is to your target times her/his guess; and it makes the number s/he earns larger, the closer her/his guess is to her/his target times your guess. Only the distance matters, NOT whether the difference is positive or negative. You earn the same number of points when your guess is too high by a given amount as when it is too low by the same amount. To pass the Understanding Test and TO BE ABLE to participate in the experiment, it is important to understand how your (respectively, her/his) original guesses should be chosen to stay within your (her/his) limits. This will be done as follows. If your original guess is below your lower limit, then your guess should be adjusted UP to your LOWER limit; and if your original guess is above your upper limit, then your guess should be adjusted DOWN to your UPPER limit. If, for example, your lower limit is 400 and you original guess 300, then your adjusted guess is 400. If your upper limit is original 600 and your guess 900, then your adjusted guess is 600. Her/his guesses are adjusted up or down to her/his lower or upper limits in the same way, except that her/his limits may be different. Remember you and s/he can choose any number within your and her/his limits, respectively. We are now starting a new round of the game. Round 1: For this round your lower limit is *$\text{LL}_{\text{you}}^{\text{Round} 1}$*, your upper limit is *$\text{UL}_{\text{you}}^{\text{Round} 1}$*, and your target is *$\text{target}_{\text{you}}^{\text{Round} 1}$*. Her/His lower limit is *$\text{LL}_{\text{his/her}}^{\text{Round} 1}$*, her/his upper limit is *$\text{UL}_{\text{his/her}}^{\text{Round} 1}$*, and her/his target is *$\text{target}_{\text{his/her}}^{\text{Round} 1}$*. Please choose a number between your lower limit and upper limit.
\end{Box1}

Similar to the repeated pBCG, we keep the whole history for each LLM subject in subsequent rounds by adding to the vector of messages (i.e., stacking messages) as the input to the API. After the user prompt from the first round, the answer given by the LLM is added to the messages in the assistant role. Different from the pBCG, and consistent with \cite{costa-gomes_cognition_2006}, we do no provide feedback to avoid repeated game effects. Therefore, after the assistant message, as a user role message, we prompt with the next round 

\begin{Box1}[]{User}
We are now starting a new round of the game. Round 2: For this round your lower limit is *$\text{LL}_{\text{you}}^{\text{Round} 2}$*, your upper limit is *$\text{UL}_{\text{you}}^{\text{Round} 2}$*, and your target is *$\text{target}_{\text{you}}^{\text{Round} 2}$*. Her/His lower limit is *$\text{LL}_{\text{his/her}}^{\text{Round} 2}$*, her/his upper limit is *$\text{UL}_{\text{his/her}}^{\text{Round} 2}$*, and her/his target is *$\text{target}_{\text{his/her}}^{\text{Round} 2}$*. Please choose a number between your lower limit and upper limit.
\end{Box1}
\noindent We repeat this process of messages, keeping the entire history of messages for each LLM subject until all 16 rounds are completed. Note that the history given to each LLM subject is specific to that LLM subject; no other information is shared with other LLM subjects beyond their history. In this way, the system message, and the rules of the game are always contained in the first 2 messages of the history, as well as the entire history of choices that each LLM subject makes.

\subsubsection{MRG}
The instructions come directly from \cite{arad_1120_2012}, with minimal modifications. Similarly, we avoid monetary payoffs as rewards to avoid issues with incentives. We also provide examples in the instructions as LLMs had better answers to the best responses understand questions (as this helps LLMs understand what kind of calculations are involved).
The system prompt is 

\begin{Box1}[]{System}
I want you to act as a human decision maker, You will be participating in 1 round of a game. You should use your best judgment to come up with responses that you like most. Always write your final answer in [ ].
\end{Box1}
\noindent This is identical for both MRG and MRG (Game 3). The following prompts include all the details for MRG and MRG (Game 3), specified by the user role.

\begin{Box1}[]{User for MRG}
You are randomly matched to play a game against a human decision maker. In the game, each of you chooses an integer between 11 and 20. Each participant will receive points in the amount they have chosen. A participant will receive an additional 20 points if they choose a number exactly one less chosen by the other player. For example, if the other participant chooses X, you will get (X-1)+20 points if you choose X-1 points. For example, if your opponent chooses 14, you will get 33 points if you choose 13. Or, if the other participant chooses 11, you will get 20 points if you choose 20. You will receive your points without knowing against whom you played. Now let's start this game. What number do you choose?
\end{Box1}
\medskip

\begin{Box1}[]{User for MRG (Game 3)}
You are randomly matched to play a game against a human decision maker. In the game, each of you chooses an integer between 11 and 20. If a participant chooses the number 20, they will receive 20 points (regardless of the other player’s choice). Otherwise, if a participant chooses any other number in this range, they will receive three points less than in the case where they choose 20. However, they will receive an additional amount of 20 points if they choose a number that is lower by exactly one than the number chosen by the other player. For example, if the other participant chooses X, you will get 17+20 points if you choose X-1. For example, if your opponent chooses 14, you will get 17+20 points if you choose 13. Or, if the other participant chooses 11 points, you will get 20 points if you choose 20. You will receive your points without knowing against whom you played. Now let's start this game. What number do you choose?
\end{Box1}

\subsection{Estimation Details}

\subsubsection{pBCG}\label{est:pBCG} 
From the theoretical predictions, we apply standard practice maximum likelihood estimation (MLE) to estimate parameters for both the level-$k$ model and the CH model. For the level-$k$ model, we estimate the proportion of responses that align with the $L_0$, $L_1$, $L_2$, $L_3$, $L_4$, and $L_\infty$ (Nash) predictions. $L_0$ also accommodates behavior which is not precisely predicted by the level-$k$ (or CH model) for levels (steps) higher than $L_1$ (step-1), i.e., the $L_0$ type captures all responses from a type who chooses uniformly randomly from the integers 0 to 100, but is not in the proximity of any prediction from any other higher type. This allows us to accommodate behavior which is not in the neighborhood of the predictions of any of the level-$k$ types (up to a finite $k$).\footnote{The goal of MLE is to find the parameter values that maximize the product of the individual likelihoods, based on a (probabilistic) matching of the theoretical predictions and the observed strategies. However, an action predicted to occur with zero probability results in a likelihood contribution of zero, rendering the overall likelihood function zero. Such an event, even if it occurs for only one observation, can obscure the predictive accuracy of any model. To avoid the zero-likelihood problem, we introduce a type that randomizes across all possible strategies, ensuring that no action is chosen with zero probability, as described in \cite{moffatt_experimetrics_2015} and footnote 2 of \cite{camerer_cognitive_2004}.} For each type, we assume that a choice is the best guess limited by their type, plus an independent and identically distributed (i.i.d) error term. Specifically, assuming the error term is i.i.d binomially distributed, the observed choice of a level-$k$ type by an individual, $i$, is given by $Y_i = l_k + \epsilon_i$, with $\epsilon_i \sim \mathcal{B}(\alpha,\beta)$ such that $\E[\epsilon_i]=0$, and $\Var[\epsilon_i]=\sigma^2$. With $\epsilon_i \in \mathbb{R}$ and $\epsilon_i = Y_i - l_k$, this allows to define a density for each of these types: 
\begin{align*}
h_k(\epsilon_i) =
    \begin{dcases}
        \binom{\alpha}{\epsilon_i}\beta^{\epsilon_i}(1-\beta)^{\alpha - {\epsilon_i}} & k\in\{1,...,K,\infty\}, \\[1em]
         \hfil \dfrac{1}{101} & k=0.
    \end{dcases}
\end{align*}

Let $\mathcal{K}$ denote the space of types, i.e., $\mathcal{K} = \{0,...,K,\infty\}$. Define proportions of types by $\mathcal{F}= \{f_{k}\}_{k\in \mathcal{K}}$, where $f_k$ is the proportion of level-$k$ types. Defining our complete parameter space as $\Theta = (\mathcal{F}, \alpha, \beta$) with $\alpha, \beta$ being the nuisance parameters, the log-likelihood function for a random sample of guesses, $\mathbf{Y} = Y_{i\in\{1,...,N\}}$ is:
\begin{equation*}
\log\mathcal{L}(\Theta|\mathbf{Y}) = \sum_{i=1}^{n} \ln \left( \sum_{k \in \mathcal{K}} f_k \cdot h_k(\epsilon_i) \right).
\end{equation*}

The purpose of assuming a binomially distributed random error is to allow for discrete-valued errors, in essence, we use a \textit{discretized} normal distribution centered around 0 (with its limit being a normal distribution). We estimate level-$k$ model with 6 levels of reasoning by setting $K = 4$ with the best guess for the $L_\infty$ type being the Nash equilibrium, the $L_0$ type being a uniform randomizer, and the remaining best guesses being determined by the theoretical predictions above.

For the CH model, $\tau$ is estimated, which represents the mean (and variance) of the frequency distribution of step-$k$ types. $\tau$ parameterizes the proportion of responses that align with step-$k$ (for $k \in \{0,...,K\}$) and step-$\infty$ predictions. The structure of the log-likelihood function is identical to the one used to estimate the level-$k$ model, with choices also modeled as noisy best guesses in the same way as above. The differences only lie in 1) the proportions of types are now determined by a Poisson distribution, i.e., the distribution of types is given by $f(k;\tau)$, and 2) the best guesses are made by assuming a distribution over lower step-$k$, determined by the conditional Poisson distribution $\{f_k(j;\tau)\}_{j=0,1,\ldots,k-1}$, i.e., with $\epsilon_i \in \mathbb{R}$ and $Y_i = s_k(\tau) + \epsilon_i$. This structure is identical to the level-$k$ model except we replace the level-$k$ best guess, $l_k$, with the CH model best guess, $s_k(\tau)$.

Let $\mathcal{K}$ denote the space of types, i.e., $\mathcal{K} = \{0,...,K,\infty\}$. Define proportions of types by $\{f(k;\tau)| k \in \mathcal{K} \}$, where $f(k;\tau)$ is the proportion of step-$k$ types, given $\tau$.\footnote{$f_{0}$ is still parameterized by $\tau$ by the law of total probability.} The complete parameter space is $\Theta = (\tau,\alpha, \beta)$ with $\alpha, \beta$ being the nuisance parameters, hence, the log-likelihood function for a random sample of guesses, $\mathbf{Y} = Y_{i\in=\{1,...,N\}}$ is:

\begin{equation*}
\log\mathcal{L}(\Theta|\mathbf{Y}) = \sum_{i=1}^{n} \ln \left( \sum_{j \in \mathcal{K}} f(\tau;k) \cdot h_k(\epsilon_i) \right).
\end{equation*}

Although the log-likelihood function is of the same structure, estimation is vastly different as the CH model is fully determined by the value of $\tau$, i.e., finding the value of $\tau$ that maximizes the log-likelihood. For consistency, we estimate using the same number of step-$k$ type players as there are level-$k$ type players.\footnote{Note that $\alpha$ and $\beta$ are simply noise parameters for the estimation in the binomial distribution, with no mapping to the theoretical models, as such, they are nuisance parameters.} 

\subsubsection{GG}\label{est:gg} 
Following the theoretical predictions, we use similar MLE techniques to estimate parameters for both the level-$k$ model and the CH model. The notable difference here is that we collect data sequentially for the 16 rounds of the GG. For each LLM, we collect these set of responses 100 times (i.e., for each LLM, there are 100 LLM subjects playing 16 rounds of the GG sequentially). With these data, for the level-$k$ model, we are able to estimate the proportion of responses that align with the $L_0$, $L_1$, $L_2$, $L_3$, $L_4$, and $L_\infty$ (Nash) predictions on an individual-response level. We do this because each round has its own different set of predictions (level-$k$, CH, Nash), so we need not base the estimation on aggregate responses, as done in the pBCG.\footnote{Naturally, another way to approach the estimation is to aggregate the data for each round, and estimate the model for each of the 16 rounds of the GG, as if each round is their own one-shot game. However, this does not make use of the sequential nature in which the GG is played in \cite{costa-gomes_cognition_2006}, which we attempt to stay as close to as possible, in terms of experimental design.} 

The observed (noisy) choice of a level-$k$ type by an individual, $i$, for each round, $r$, is given by $Y_i^r = l_k^r + \epsilon_i^r$, with $\epsilon_i^r \sim \mathcal{B}(\alpha^r,\beta^r)$ such that $\E[\epsilon_i^r]=0$, and $\Var[\epsilon_i^r]={(\sigma^r)}^2$. With $\epsilon_i^r \in \mathbb{R}$ and $\epsilon_i^r = Y_i^r - l_k^r$, this allows to define a density for each of these types: 

\begin{align*}
h_k^r(\epsilon_i) =
    \begin{dcases}
        \binom{\alpha^r}{\epsilon_i^r}\beta^{\epsilon_i^r}(1-\beta)^{\alpha^r - {\epsilon_i^r}} & k\in\{1,...,K,\infty\}, \\[1em]
         \hfil \dfrac{1}{b_i^r - a_i^r} & k=0.
    \end{dcases}
\end{align*}

Depending on the parameters of the GG, the level of reasoning required to reach the $L_\infty$ prediction can be fewer than $K(>1)$. In these cases, we assume the response is only consistent with the $L_\infty$ (and $L_0$) type. Formally, for any $k\in\{1,...,K\}$ and $l_{k}^r = l_\infty^r$, set $h_j^r(\epsilon_i)=0$, for $j\leq k$, with $h_\infty^r(\epsilon_i)$ and $h_0^r(\epsilon_i)$ defined as above.\footnote{While the majority of the mass goes to the $L_\infty$ proportion when the response is close to the $L_\infty$ prediction, we still allow for random behavior.}

Let $\mathcal{K}$ denote the space of types, i.e., $\mathcal{K} = \{0,...,K,\infty\}$. Define proportions of types by $\mathcal{F}= \{f_{k}\}_{k\in \mathcal{K}}$, where $f_k$ is the proportion of level-$k$ types. Defining our complete parameter space as $\Theta = (\mathcal{F}, \bm{\alpha}, \bm{\beta}$) with $\bm{\alpha}, \bm{\beta}$ collecting nuisance parameters in vectors, the log-likelihood function for a random sample of guesses, $\mathbf{Y}_i = Y_{i,r\in\{1,...,16\}}$ is:
\begin{equation*}
\log\mathcal{L}(\Theta|\mathbf{Y}_i) = \sum_{r=1}^{16} \ln \left( \sum_{k \in \mathcal{K}} f_k \cdot h_k^r(\epsilon_i) \right).
\end{equation*}

Notice this is virtually identical to the MLE used in the pBCG except we use data collected for all rounds per LLM subject to estimate the level-$k$ model, rather than aggregating over all the data collected. In this way, we can estimate the level-$k$ model for each LLM subject. As we estimate up to $L_4$, including $L_0$ and $L_\infty$, we set $K=4$. 

For the CH model, $\tau$ is estimated in an analogous way. Define proportions of types in each round by $\{f(k;\tau)| k \in \{0,...,K,\infty\} \}$, where $f(k;\tau)$ is the proportion of step-$k$ types, given $\tau$, with $\epsilon_i^r \in \mathbb{R}$ and $Y_i^r = s_k^r(\tau) + \epsilon_i^r$. As before we impose that, for any $k\in\{1,...,K\}$ and $l_{k}^r = l_\infty^r$, set $h_j^r(\epsilon_i)=0$, for $j\leq k$, with $h_\infty^r(\epsilon_i)$ and $h_0^r(\epsilon_i)$ defined as above. The complete parameter space is $\Theta = (\tau,\bm{\alpha}, \bm{\beta})$ with $\bm{\alpha}, \bm{\beta}$ collecting nuisance parameters in vectors, hence, the log-likelihood function for a random sample of guesses, $\mathbf{Y}_i = Y_{i,r\in\{1,...,16\}}$ is:

\begin{equation*}
\log\mathcal{L}(\Theta|\mathbf{Y}_i) = \sum_{r=1}^{16} \ln \left( \sum_{k \in \mathcal{K}} f(\tau;k) \cdot h_k^r(\epsilon_i) \right).
\end{equation*}

Again, the log-likelihood function is of the same structure with vastly different estimation as the CH model is fully determined by $\tau$, separately for each LLM subject. For consistency, we estimate using the same number of step-$k$ type players as there are level-$k$ type players.\footnote{Note that $\bm{\alpha}$ and $\bm{\beta}$ are again (vectors of) noise parameters for the estimation in the binomial distribution, with no mapping to the theoretical models.} 

\subsubsection{MRG}\label{est:mrg} 
\noindent\textbf{Estimation.} Similarly, from the theoretical predictions, we apply MLE to estimate parameters for both the level-$k$ model and the CH model analogously to what was done for the pBCG and GG. We keep the same number of levels/steps as the pBCG but replace $L_\infty$ with $L_4$, as there is no pure Nash equilibrium. We still accommodate behavior which is not predicted by the level-$k$ or CH model by including a purely random type. However, as $L_0$ behavior can be very distinct from purely random behavior, we specifically distinguish between these two types. Another reason for doing this is to allow for the uniform randomization choice (15), and lower choices, to be considered as random behavior, rather than a very high stage of reasoning.\footnote{This is consistent with the empirical results in \cite{arad_1120_2012} with virtually no subject choosing lower than 15.}  Estimation is done separately for each of MRG, and MRG (Game 3). 

Given the finite nature of the actions in MRG and MRG (Game 3), applying MLE to estimate the parameters for the level-$k$ model and the CH model is simple and standard. For the level-$k$ model, we estimate the proportion of levels that are consistent with purely random behavior, $L_0$, $L_1$, $L_2$, $L_3$, and $L_4$ predictions. Let $\mathcal{K}$ denote the space of types, i.e., $\mathcal{K}=\{\text{random},0,1,...,K\}$. We define the density for each type as follows:

\begin{align*}
h_k(Y_i) =
    \begin{dcases}
         \mathbbm{1}[Y_i = l_k ] & k\in\{0,...,K\}, \\[1em]
         \hfil \dfrac{1}{10} & k=\text{\{random\}}.
    \end{dcases}
\end{align*}

where $\mathbbm{1}[Y_i = l_k ]$ is an indicator function which takes 1 when $Y_i = l_k $ and 0 otherwise. Define proportions of types by $\mathcal{F} = \{ f_k\}_{k \in \mathcal{K}}$, where $f_k$ is the proportion of each type. We set $K=4$ to estimate 6 levels of reasoning (including $L_0$, and purely random type). The parameter space is simply $\Theta = \mathcal{F}$ and the log-likelihood function for a random sample of guesses, $\mathbf{Y} = Y_{i\in\{1,...,N\}}$ is:
\begin{equation*}
\log\mathcal{L}(\Theta|\mathbf{Y}) = \sum_{i=1}^{n} \ln \left( \sum_{k \in \mathcal{K}} f_k \cdot h_k(Y_i) \right).
\end{equation*}

For the CH model, we estimate $\tau$ to parameterize the proportion of responses that align with step-$k$, for $k\in\{\text{random},0,1,...,K\}$, to be consistent with the level-$k$ model. As before, the proportions of types are determined by a Poisson distribution with best responses as given above in the theoretical predictions. Define proportions of types by $\{ f(k;\tau) | k \in \mathcal{K} \}$ where $f(k;\tau)$ is the proportion of step-$k$ types, given $\tau$. The parameter space is simply $\Theta = \{\tau\} $ and the log-likelihood function for a random sample of guesses, $\mathbf{Y} = Y_{i\in\{1,...,N\}}$ is:
\begin{equation*}
\log\mathcal{L}(\Theta|\mathbf{Y}) = \sum_{i=1}^{n} \ln \left( \sum_{k \in \mathcal{K}} f(k;\tau) \cdot \mathbbm{1}[Y_i = s_k(\tau) ] \right).
\end{equation*}

As with the pBCG and GG, the log-likelihood function is of the same structure between models, with different estimation, given the way that the proportions are parameterized. For consistency, the same number of step-$k$ players is used as there are level-$k$ players. 

\subsection{Temperature}\label{sec:temp}

The temperature parameter influences outputs in terms of randomness and creativity. A lower temperature value results in more predictable and deterministic responses, in favor of words with the highest probabilities. Conversely, a higher temperature results in more diverse and novel responses by increasing the chance of less likely words. In line with studies exploring the impact of temperature settings on LLM behavior \citep{loya_exploring_2023,binz_using_2023,zhu_hot_2024}, every condition mentioned is also tested at 3 temperature levels. The baseline is set at a moderate temperature of 0.5 to ensure the most balanced approach. We have higher (0.75) and lower (0.25) temperature level conditions to examine how variations in randomness influence (observed) strategic sophistication. The exception is with \gptoone as at the time of collecting the data (October 2024) varying the temperature parameter was not available. Given the results of the baseline (for all games) and as we only use \gptoonens, \claudethreens, and \geminitwo for the GG, we did not run the GG with different temperatures. Given the results from the level-$k$ model have very little variation across temperature, the CH model estimates also do not vary much from baseline temperature either. 
\subsubsection{pBCG}

For all conditions (baseline and alternate), we collect data varying the temperature to determine whether there is any impact on LLM responses. Overall, the results remained largely consistent with those observed at the baseline temperature. The only differences of note occurred in the higher temperature condition, in which there was an increase in incoherent answers, and a slight rise in the variation of responses.\footnote{We still maintain the planned sample size after dropping incoherent answers.} Analogously, for lower temperatures, the modal responses became somewhat more modal, as expected with more deterministic results.  However, the estimates of either model remain relatively unchanged. This consistency across different temperatures suggests that strategic sophistication of LLMs is robust to changes in temperature. We present the results for the different temperature conditions below.

\begin{figure}[H]
    \centering
    \includegraphics[width=0.8\textwidth]{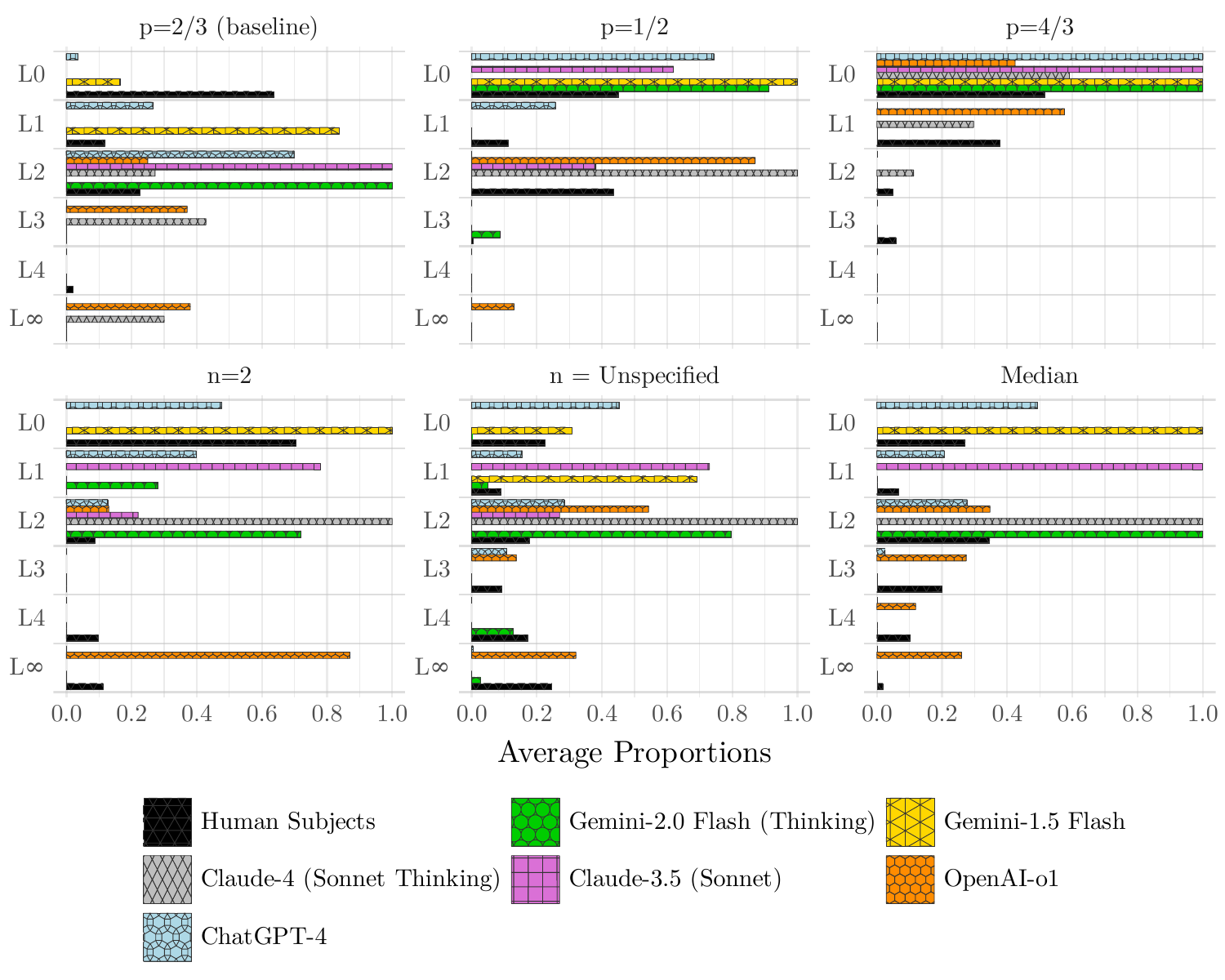}
    \caption{Estimates of the proportions of level-$k$ types for the pBCG with high temperature.}
    \label{fig:baseline_lk_temphigh}
\end{figure}

\begin{figure}[H]
    \centering
    \includegraphics[width=0.8\textwidth]{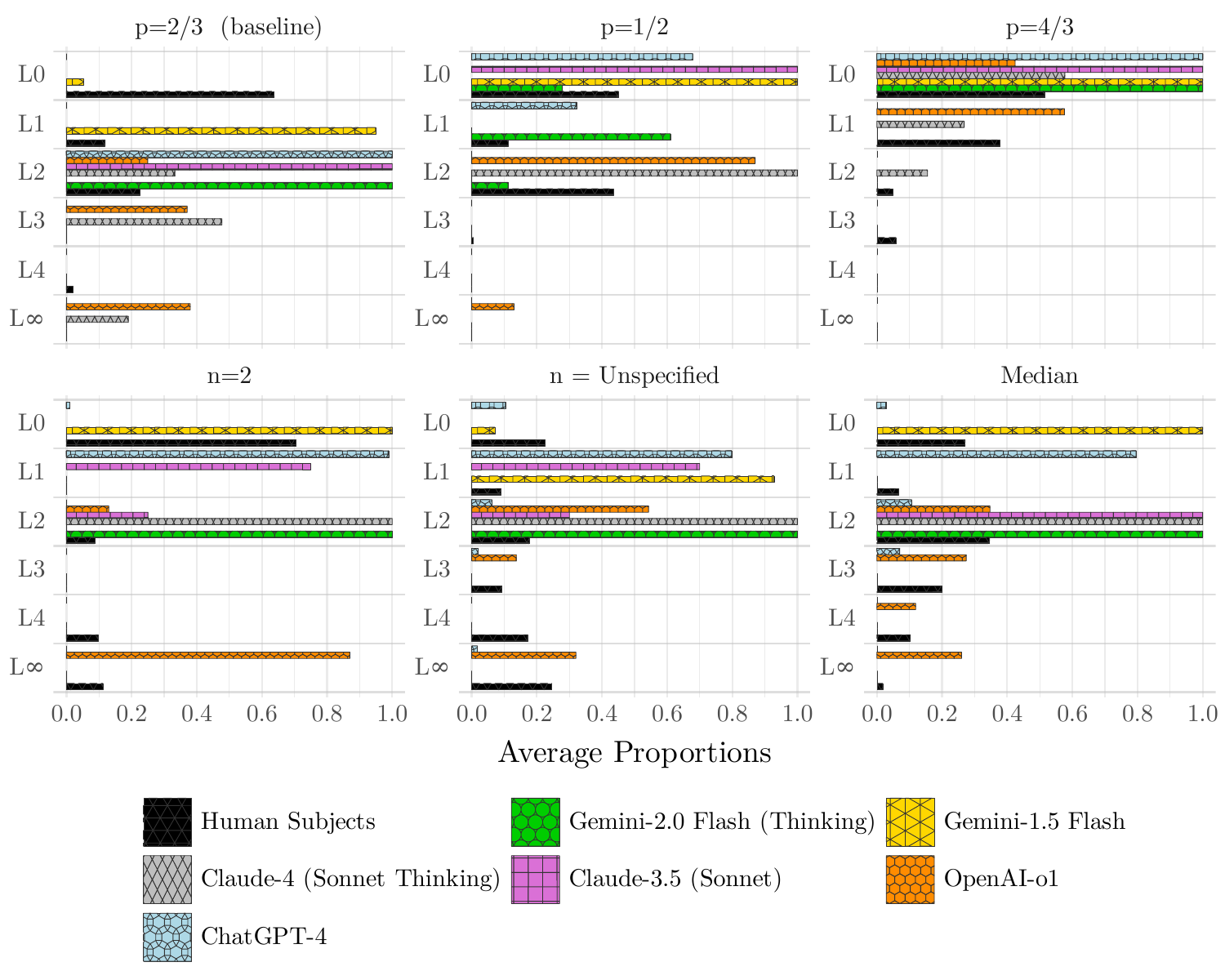}
    \caption{Estimates of the proportions of level-$k$ types for the pBCG with low temperature.}
    \label{fig:baseline_lk_templow}
\end{figure}

\begin{figure}[H]
\centering
    \includegraphics[width=\linewidth]{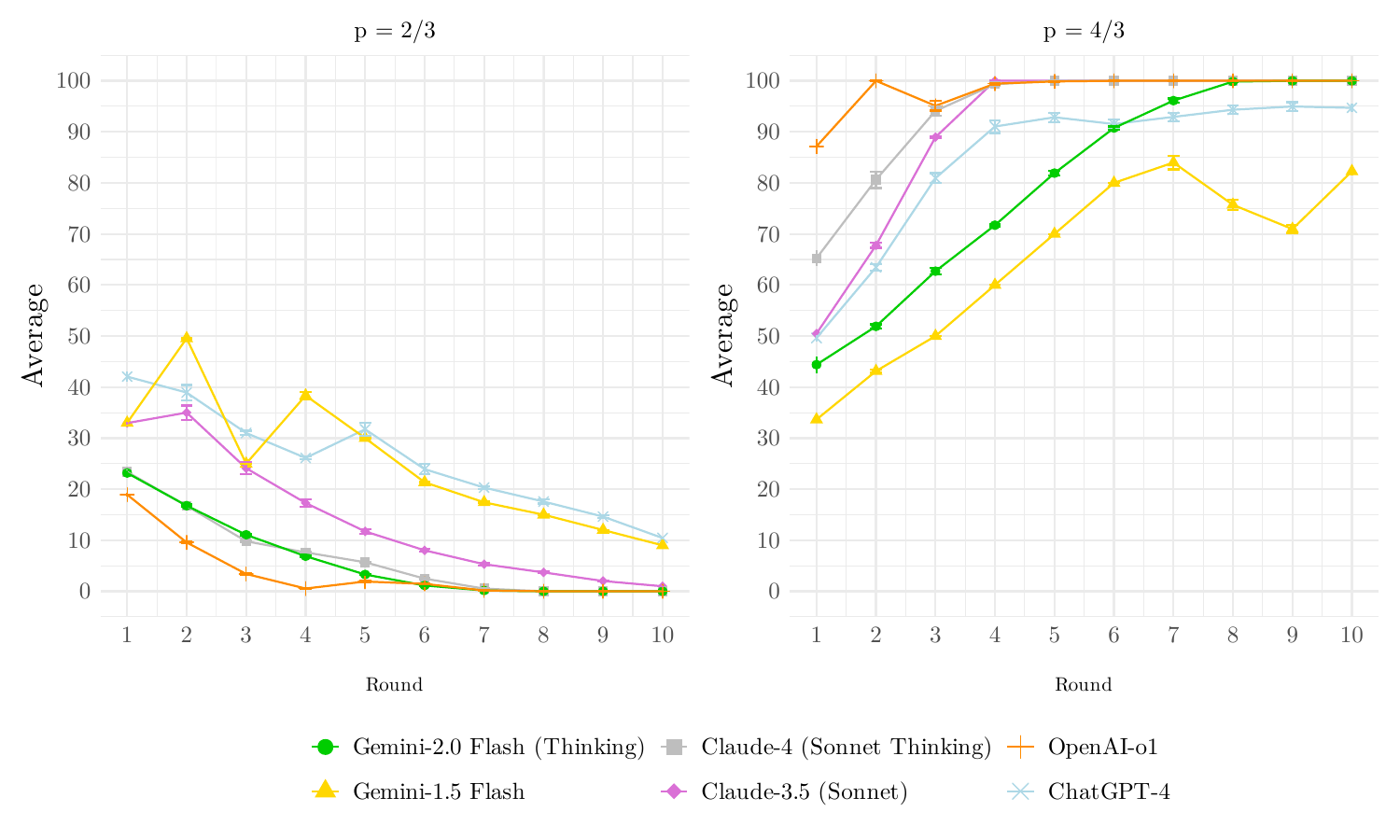}
    \caption{Time series of responses for multiple rounds pBCG with high temperature. This figure consists of two figures for p = 2/3 and p = 4/3, respectively. Each figure shows the average responses in each round for each LLM. Vertical bars are 95\% confidence intervals}
    \label{fig:feedback_temphigh}
\end{figure}

\begin{figure}[H]
\centering
    \includegraphics[width=\linewidth]{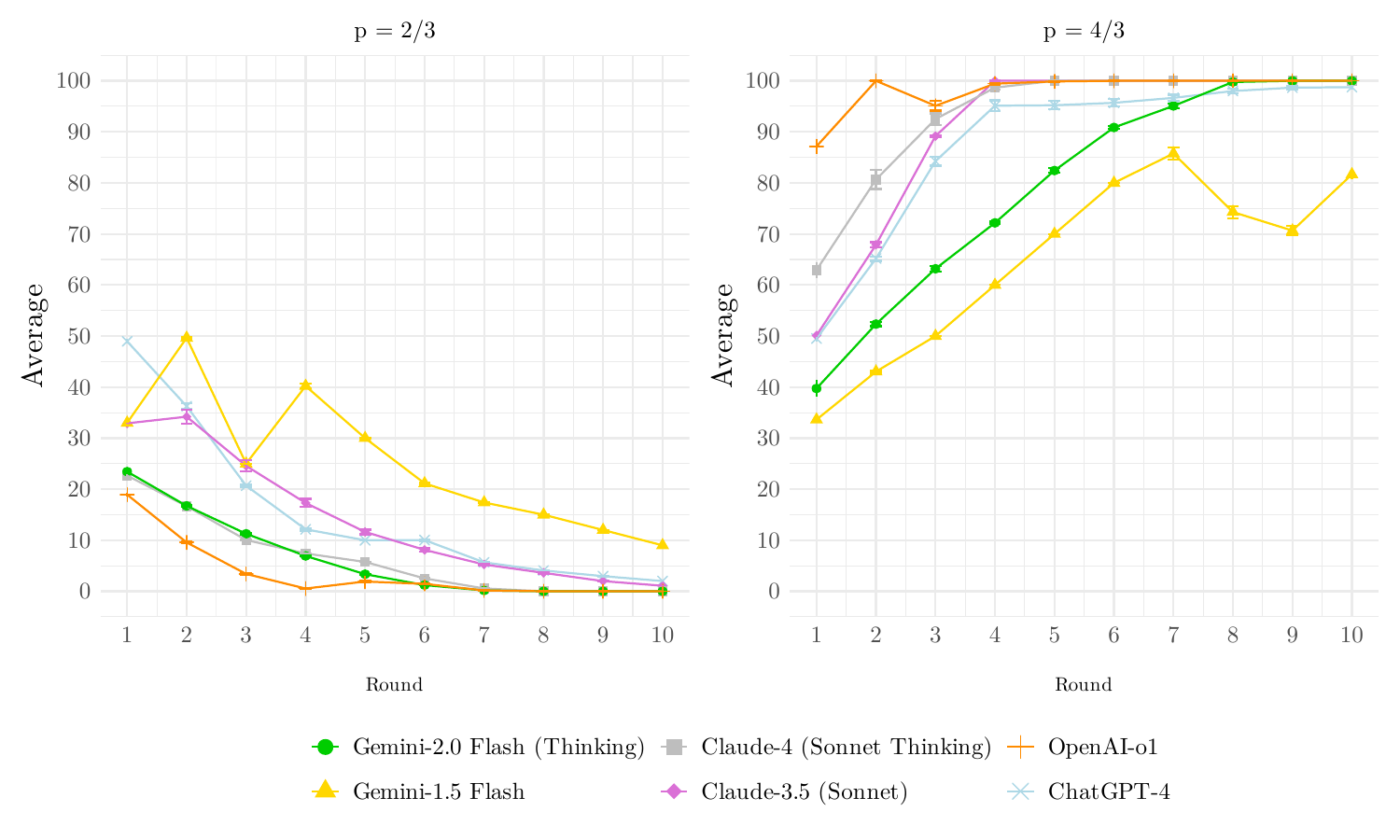}
    \caption{Time series of responses for multiple rounds pBCG with low temperature. This figure consists of two figures for p = 2/3 and p = 4/3, respectively. Each figure shows the average responses in each round for each LLM. Vertical bars are 95\% confidence intervals}
    \label{fig:feedback_templow}
\end{figure}

\subsubsection{MRG}

\begin{figure}[H]
    \centering
    \includegraphics[width=\textwidth]{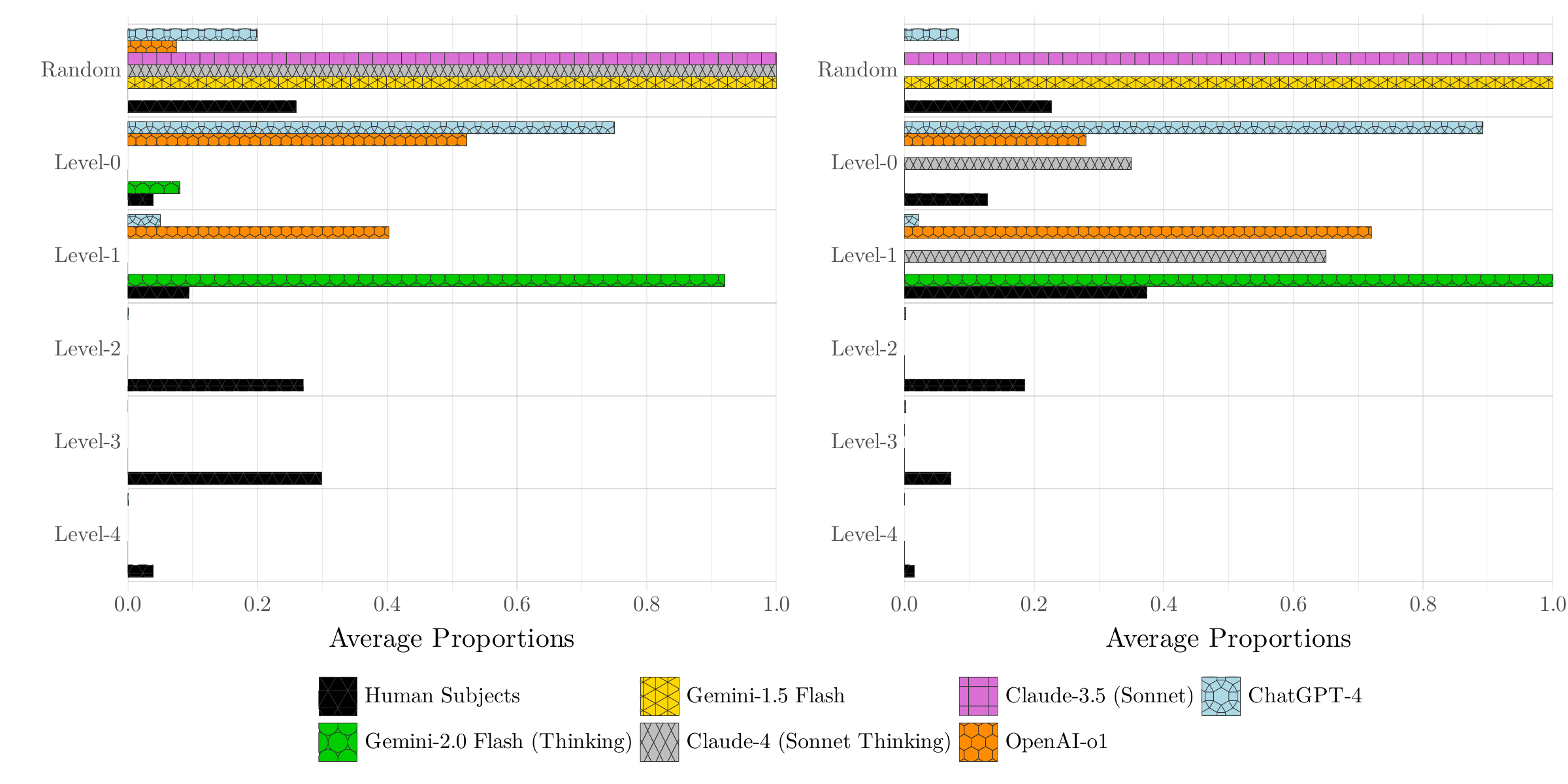}
    \caption{Estimates of the proportions of level-$k$ types for MRG, and MRG (Game 3) with high temperature.}
    \label{fig:mrg_lk_hightemp}
\end{figure}

\begin{figure}[H]
    \centering
    \includegraphics[width=\textwidth]{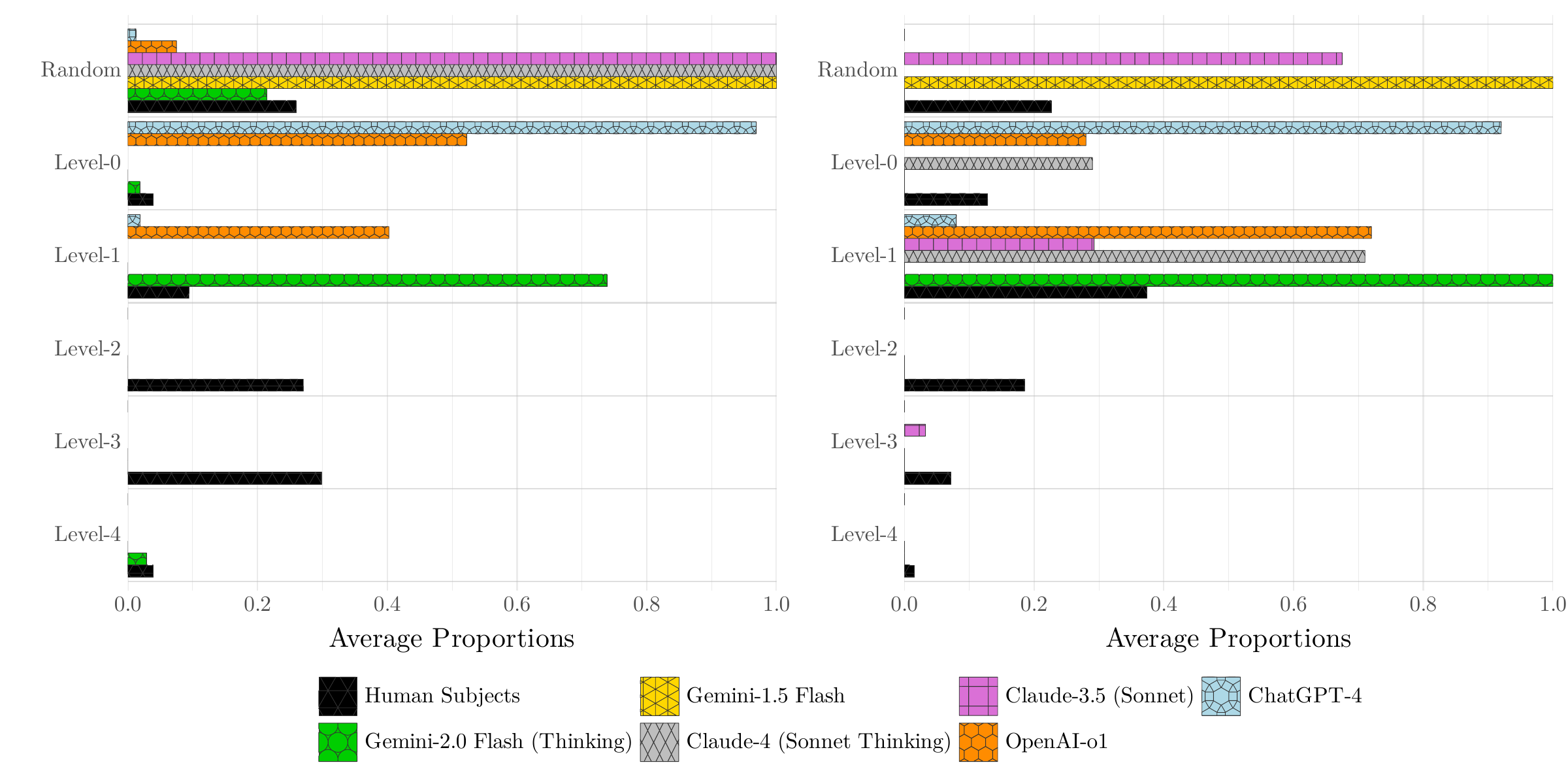}
    \caption{Estimates of the proportions of level-$k$ types for MRG, and MRG (Game 3) with low temperature.}
    \label{fig:mrg_lk_lowtemp}
\end{figure}

\end{document}